\documentclass[prb,showpacs,preprintnumbers,amsmath,superscriptaddress,twocolumn]{revtex4} %
\usepackage{mathrsfs}

\usepackage{graphicx}
\usepackage{dcolumn}
\usepackage{bm}
\usepackage{amsmath}

\makeatletter
\renewcommand{\section}{\@startsection{section}{1}{0mm}
  {-\baselineskip}{0.5\baselineskip}{\bf\leftline}}
\makeatother

\makeatletter
\renewcommand{\subsection}{\@startsection{subsection}{1}{0mm}
  {-\baselineskip}{0.5\baselineskip}{\bf\leftline}}
\makeatother

\makeatletter
\renewcommand{\subsubsection}{\@startsection{subsubsection}{1}{0mm}
  {-\baselineskip}{0.5\baselineskip}{\bf\;\leftline}}%
\makeatother

\begin{document}

\preprint{APS/123-QED}
\title{Transferable tight binding model for strained group IV and III-V materials and heterostructures}

\author{Yaohua~Tan}
\email{tyhua02@gmail.com}
\affiliation{%
School of Electrical and Computer Engineering,Network for
Computational Nanotechnology,Purdue University, West Lafayette,
Indiana 47906, USA 
}
\affiliation{%
Department of Electrical and Computer Engineering, University of Virginia, Charlottesville,
Virginia 22904, USA
}
\author{Michael~Povolotskyi}
\affiliation{%
School of Electrical and Computer Engineering,Network for
Computational Nanotechnology,Purdue University, West Lafayette,
Indiana 47906, USA 
}
\author{Tillmann~Kubis }
\affiliation{%
School of Electrical and Computer Engineering,Network for
Computational Nanotechnology,Purdue University, West Lafayette,
Indiana 47906, USA 
}
\author{Timothy B.~Boykin} \affiliation{
University of Alabama in Huntsville,Huntsville, Alabama 35899, USA 
}%
\author{Gerhard~Klimeck }
\affiliation{%
School of Electrical and Computer Engineering,Network for
Computational Nanotechnology,Purdue University, West Lafayette,
Indiana 47906, USA  
}
%
%
%

%
%

\date{\today}
\begin{abstract}
It is critical to capture the effect due to strain and material
interface for device level transistor modeling. We introduced a
transferable sp3d5s* tight binding model with nearest neighbor interactions
for arbitrarily strained group IV and III-V materials. 
The tight binding model is parameterized with respect to Hybrid functional(HSE06) calculations for varieties of strained systems. The tight binding calculations of ultra small superlattices formed by group IV and
group III-V materials show good agreement with the corresponding
HSE06 calculations. The application of tight binding model to
superlattices demonstrates that transferable tight binding model with nearest neighbor interactions can be obtained for group IV and III-V materials.
\end{abstract}

\maketitle

\section{Introduction }
Modern field effect transistors have reached critical device
dimensions in sub-10 nanometer. To surpass the coming limits of
downscaling of field effect transistor, innovative devices such as
tunneling field-effect
transistors(TFET)~\cite{TFET_King,TFET_CNT,TFET_GNR} and
superlattice field-effect
transistors~\cite{Superlattice_Fets_original,Superlattice_fets_PengyuLong}
are actively investigated. Those devices rely strongly on the usage
of hetero-structures and strain techniques. To have reliable
prediction of the performance in those devices, it is critical to
have a atomistic model that is able to model strained ultra-small
heterostructures accurately.

\textit{Ab-initio} methods offer atomistic representations with
subatomic resolution for a variety of materials and
heterostructures. However, accurate \textit{ab-initio} methods, such
as Hybrid functionals\cite{HSE06_original,HSE06},
GW\cite{Hedin_GW,Hybertsen_GW} and BSE\cite{BSE_PRL} approximations
are in general computationally too expensive to be applied to
systems with a size of realistic device. Furthermore, those methods
assume equilibrium and cannot truly model out-of-equilibrium device
conditions where e.g. a large voltage might have been applied to
drive carriers. For these reasons, more efficient semi-empirical
approaches, such as the
$\textrm{k}\cdot\textrm{p}$\cite{8bandkdotp_Bahder,YaohuaTan_kp_strain,Jun_NEGF_kp},
the empirical pseudopotential\cite{Fischette_PseudoPotential} and
the empirical tight-binding(ETB)
methods\cite{Jancu_Tightbinding,Boykin_TB_strain,YaohuaTan_abinitioMapping} are actively
developed.

Among these empirical approaches, ETB method has established itself
as the standard state-of-the-art basis for realistic device
simulations\cite{Nemo5_2013}. ETB has been successfully applied to
electronic structures of millions of atoms~\cite{Klimeck_QuantumDot}
as well as on non-equilibrium transport problems that even involve
inelastic scattering~\cite{Klimeck_RTD}. For strained systems,
modified ETB models take into account the altered environment in
terms of both bond angle and length. In the simplest tight binding
strain model, generalized Harrison's law\cite{Harrison,Jancu_Tightbinding,
SiGe_harrison'slaw} is usually adopted to describe bond-length
dependence of the nearest-neighbor coupling parameters. Changes of
bond angles in interatomic interactions are automatically
incorporated through the Slater-Koster
formulas\cite{Podolskiy_TBElements}. This simplest tight binding
strain model can reproduce some hydrostatic and uniaxial deformation
potentials\cite{Jancu_Tightbinding}, while much higher accuracy can
be achieved by introducing the strain-dependent onsite parameters.
Boykin et al. \cite{Boykin_TB_strain} introduced nearest neighbor
position dependent diagonal orbital energies to the sp3d5s* tight
binding model to reproduce correct deformations under [001] strains.
Off-diagonal onsite corrections are suggested by Niquet et al.
\cite{Niquet_TB_model} and Boykin et al.\cite{Boykin2010} to model
the strain behavior of indirect conduction valleys of materials with diamond structures under [110] strains.
\begin{figure}[h]
\includegraphics[width=0.4\textwidth]{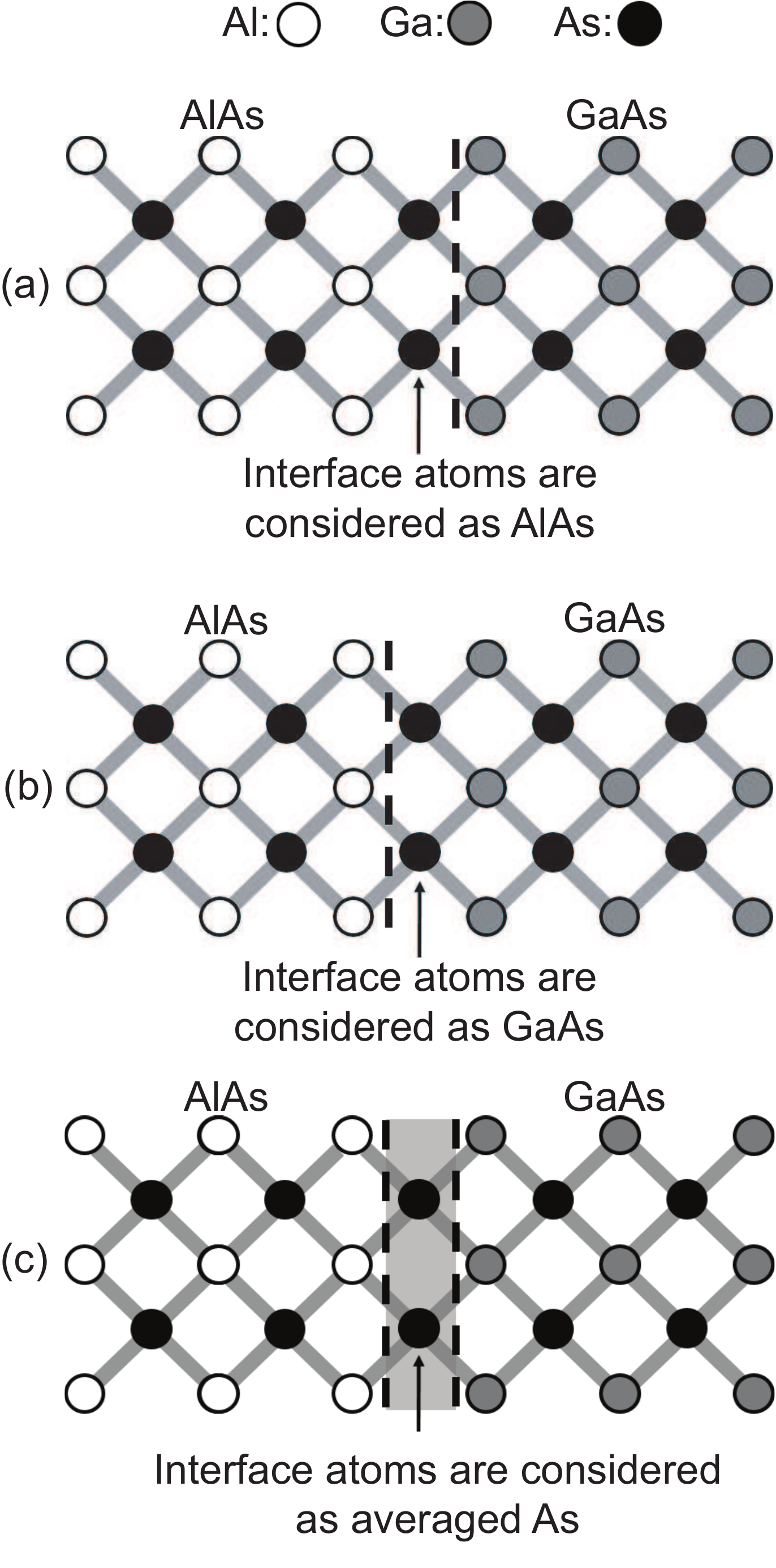}
\caption{Different definitions of materials at GaAs/AlAs interfaces.
Regions of AlAs and GaAs materials are separated by the dashed line.
In all presented definitions, the left parts are AlAs and the right parts
are GaAs. In definition (a), the interface As atoms are defined as
atoms in AlAs; in definition (b), the interface As atoms are defined
as atoms in GaAs. In the case (c), the interface As atoms are
defined as As atoms in an averaged material of AlAs and GaAs. }
\label{fig_material_def}
\end{figure}

Those existing ETB strain models are fitted to pure strained bulk material
instead of more complicated nanostructures. However, the transferability of those ETB models and parameters is questionable when applied to heterostructures. 
First of all, traditional ETB parameters depend on material types,
while material type around interfaces can not be clearly defined.
Fig.\ref{fig_material_def} shows three possible definitions of
materials near a GaAs/AlAs interface. Interface As atoms are
interpreted as atoms in either (a) As of AlAs or (b) As of GaAs.
Another usual assumption, shown by definition (c),is to take the
interface As atoms to have an average of the onsite potentials. All
those definitions are customarily used but with no hard data to
justify. Secondly, it was shown that ETB parameters obtained by
direct fitting possibly lead to unphysical results in
nano-structures like ultra-thin
bodies\cite{YaohuaTan_abinitioMapping,DFTMapping_JCEL}. To improve the
transferability of ETB parameters, \textit{ab-initio} mapping
methods are developed in ref \onlinecite{YaohuaTan_abinitioMapping}.
This method is an \textit{ab-initio} wave functions based tight
binding parameterization algorithm. With this method, it is shown
that ETB models are transferable to Si and GaAs ultra thin bodies.

In this paper, a new ETB model for strained materials considering only nearest neighbor interactions is introduced for strained group IV and III-V
semiconductors. This strain model takes account of arbitrary strain
effects to band structure. Transferable ETB parameters for strained
III-Vs and group IV materials are obtained by \textit{ab-initio}
mapping algorithm from Hybrid functional calculations. The ETB model
shows good transferability when applied to strained superlattices.

This paper is organized as follows. In section~\ref{sec:method}, the
ETB model for strained materials is described.
Section~\ref{sec:results} shows the validation of the ETB model for
strained systems and superlattices. Subsection~\ref{sec:parameters}
describes the details of getting ETB parameters; ETB parameters for
strained group IV and III-V materials are listed in this section.
Subsection~\ref{sec:strained_materials_cmp} compares the tight
binding and hybrid functional results for unstrained and strained
materials. Subsection~\ref{sec:superlattices_cmp} presents the
application of ETB model in strained superlattices, the tight
binding results for superlattices are compared with hybrid
functional calculations. Finally, the  ETB model of strained materials and
corresponding results are summarized in
Section~\ref{sec:Conclusion}.

\section{Model}\label{sec:method}
The ETB model of strained materials in this work is based on the
multipole expansion\cite{Multipole_expansion} of the local potential
near each atom. This ETB model has environment dependency, and it
does not rely on the selection of coordinates. It can be applied to
arbitrarily strained and rotated systems. In this work, the tight
binding model is applied to group IV and III-V semiconductors which have 
diamond or zincblende structures. However, the application of this model is in
principle not limited to group IV and III-V semiconductors. For materials considered in this work, the interaction range considered in the tight
binding model is limited to the first nearest neighbors. In the
following sections, letters in bold such as $\mathbf{r}$ and
$\mathbf{d}$ are used for three dimensional vectors; correspondingly, $r$ and $d$ are 
used to denote the lengths of $\mathbf{r}$ and $\mathbf{d}$.
$\Omega$ stands for polar angle and $\theta$ and azimuth angle
$\phi$ of a three dimensional vector. $\alpha$, $\beta$ and $\gamma$
correspond to tuples of angular and magnetic quantum numbers $l_1 m_1$,$l_2
m_2$ and $l_3 m_3$ of ETB orbitals respectively. Dirac notation is
used for ETB basis functions, e.g. $|\psi_{\alpha_i} \rangle$ stands
for $\alpha$ orbital of atom $i$.

\begin{table*}
  \centering
  \begin{tabular}{c|c|c|c|c|c|c|c|c}
    \hline\hline
    \begin{tabular}{c}
     Atom \\
     \hline
    $E_{s}$  \\
    $E_{p}$  \\
    $E_{s^*}$ \\
    $E_{d}$ \\
    $\Delta$ \\
    \end{tabular}
    &
\begin{tabular}{c}
  Si \\
  \hline
       1.1727 \\
       10.1115 \\
       12.4094 \\
       13.8987 \\
       0.0215 \\
\end{tabular}
    &
\begin{tabular}{c}
  Ge \\
  \hline
       -0.1105 \\
       9.8495 \\
       12.9983 \\
       13.3211 \\
       0.1234 \\
\end{tabular}
&
\begin{tabular}{c}
  Al \\
  \hline
       2.5246 \\
       8.8642 \\
       12.701 \\
       13.540 \\
       0.0015\\

\end{tabular}
&
\begin{tabular}{c}
  Ga \\
  \hline
       1.4880 \\
       8.6528 \\
       12.7318 \\
       13.5576 \\
       0.0243 \\
\end{tabular}
&
\begin{tabular}{c}
  In \\
  \hline
       1.6787 \\
       8.9987 \\
       12.7742 \\
       13.5664 \\
       0.1301 \\
\end{tabular}
&
\begin{tabular}{c}
  P \\
  \hline
       -2.3788 \\
       7.6742 \\
       12.5016 \\
       13.0781 \\
       0.0252 \\
\end{tabular}
&
\begin{tabular}{c}
  As \\
  \hline
       -3.5206 \\
       7.6037 \\
       12.5733 \\
       13.1056 \\
       0.1293 \\
\end{tabular}
&
\begin{tabular}{c}
  Sb \\
  \hline
       -2.3695 \\
       6.8994 \\
       12.6421 \\
       13.1316 \\
       0.2871 \\
\end{tabular}
    \\
  \hline
  \end{tabular}
  \caption{ Atom type dependent onsite and spin orbit coupling parameters for group IV and III-V elements.
  All parameters in this table have the unit of eV. }\label{tab:onsite_soc}
\end{table*}

\begin{table*}
  \centering
  \begin{tabular}{c|c|c|c|c|c|c|c|c|c|c|c|c}
    \hline\hline
    \begin{tabular}{c}
     bond \\
     \hline
                  $I_{s_c,a}$ \\
                  $I_{p_c,a}$ \\
              $I_{s^*_c,a}$ \\
                  $I_{d_c,a}$ \\
        	$\Delta_{ca}$ \\
            $\lambda_{s_c,a}$ \\
            $\lambda_{p_c,a}$ \\
        $\lambda_{s^*_c,a}$ \\
            $\lambda_{d_c,a}$ \\\hline
                  $I_{s_a,c}$ \\
                  $I_{p_a,c}$ \\
              $I_{s^*_a,c}$ \\
                  $I_{d_a,c}$ \\
        	$\Delta_{ac}$ \\
            $\lambda_{s_a,c}$ \\
            $\lambda_{p_a,c}$ \\
        $\lambda_{s^*_a,c}$ \\
            $\lambda_{d_a,c}$ \\
            \hline
        $O_{ac}$\\
        $\lambda_{ac}$\\ 
        \hline
        $\delta d_{ij}({\AA})$\\
    \end{tabular}
    &
\begin{tabular}{c}
  \hline
  Si-Si \\
  \hline
       3.1457 \\
       2.5307 \\
       6.7086 \\
       3.5979 \\
       0.0\\
       1.3389 \\
       1.4197 \\
       0.9522 \\
       1.1200 \\
          \hline
           - \\
           - \\
           - \\
           - \\
           - \\
           - \\
           - \\
           - \\
           - \\
            \hline
            -2.1211\\
            1.3004\\
            \hline
            -0.0118\\
\end{tabular}
    &
\begin{tabular}{c}
  \hline
  Ge-Ge \\
  \hline
       2.4312 \\
       2.0823 \\
       6.3232 \\
       3.4105 \\
       0.0\\
       1.4676 \\
       1.5634 \\
       1.0074 \\
       1.1411 \\
           \hline
           - \\
           - \\
           - \\
           - \\
           - \\
           - \\
           - \\
           - \\
           - \\
            \hline
            -1.5926\\
            1.5457\\
            \hline
            -0.0043\\
\end{tabular}
    &
\begin{tabular}{c}
  Ge-Si \\
  \hline
       2.8305 \\
       2.3208 \\
       6.5890 \\
       3.6100 \\
       0.0\\
       1.4230 \\
       1.5026 \\
       0.9195 \\
       1.2181 \\
\hline
       2.6494 \\
       2.2079 \\
       6.8296 \\
       3.2390 \\
       0.0\\
       1.4527 \\
       1.5253 \\
       0.6896 \\
       1.0880 \\
                        \hline
            -1.5267\\
            1.4566\\
            \hline
            -0.08\\
\end{tabular}
&
\begin{tabular}{c}
  Al-P \\
  \hline
       3.4070 \\
       2.8113 \\
       5.8451 \\
       3.1639 \\
       0.0023 \\
       1.3373 \\
       1.2648 \\
       0.9476 \\
       1.1719 \\
\hline
       2.1735 \\
       1.8851 \\
       5.6415 \\
       3.2039 \\
       0.0003 \\
       1.5196 \\
       1.4219 \\
       1.0031 \\
       1.0961 \\
            \hline
            -2.0783\\
            1.1878\\
            \hline
            0.0537\\
\end{tabular}
&
\begin{tabular}{c}
  Al-As \\
  \hline
       3.3106 \\
       2.7621 \\
       5.8293 \\
       3.0491 \\
       0.0 \\
       1.4021 \\
       1.3069 \\
       0.9414 \\
       1.1437 \\
       \hline
       2.2018 \\
       1.8730 \\
       5.7855 \\
       3.0758 \\
       0.0017 \\
       1.4805 \\
       1.4592 \\
       1.0470 \\
       1.2393 \\
            \hline
            -1.9981\\
            1.2167\\
            \hline
            0.0217\\
\end{tabular}
&
\begin{tabular}{c}
  Al-Sb \\
  \hline
       3.9775 \\
       3.5125 \\
       6.1929 \\
       3.4694 \\
       0.0019 \\
       1.4438 \\
       1.3917 \\
       0.8225 \\
       1.3307 \\
       \hline
       3.1709 \\
       2.9550 \\
       6.5076 \\
       3.6241 \\
       0.0045 \\
       1.5553 \\
       1.4960 \\
       1.0632 \\
       1.4126 \\
            \hline
            -2.7374\\
            1.2930\\
            \hline
            -0.0081\\
\end{tabular}
&
\begin{tabular}{c}
  Ga-P \\
  \hline
       2.5762 \\
       2.3797 \\
       5.6393 \\
       2.8883 \\
       0.0142 \\
       1.4641 \\
       1.1584 \\
       0.9348 \\
       1.1312 \\
       \hline
       1.7691 \\
       1.5278 \\
       5.5291 \\
       2.9033 \\
       0.0015 \\
       1.5613 \\
       1.6773 \\
       0.9848 \\
       1.1407 \\
            \hline
            -1.6875\\
            1.2657\\
            \hline
            0.0043\\
\end{tabular}
&
\begin{tabular}{c}
  Ga-As \\
  \hline
       2.4389 \\
       2.3491 \\
       5.6115 \\
       2.8751 \\
       0.0097 \\
       1.4410 \\
       1.2383 \\
       0.9153 \\
       1.2365 \\
       \hline
       1.8772 \\
       1.5867 \\
       5.5735 \\
       2.8658 \\
       0.0053 \\
       1.5505 \\
       1.6838 \\
       1.0037 \\
       1.0993 \\
            \hline
            -1.6467\\
            1.2697\\
            \hline
            -0.0098\\
\end{tabular}
&
\begin{tabular}{c}
  Ga-Sb \\
  \hline
       2.6906 \\
       2.7446 \\
       5.5030 \\
       3.0666 \\
       0.0003 \\
       1.5640 \\
       1.5635 \\
       1.0551 \\
       1.4301 \\
       \hline
       2.5078 \\
       2.2820 \\
       5.7372 \\
       2.8997 \\
       0.0085 \\
       1.6001 \\
       1.6629 \\
       0.8877 \\
       1.3835 \\
            \hline
            -1.9763\\
            1.2931\\
            \hline
            -0.0019\\
\end{tabular}
&
\begin{tabular}{c}
  In-P \\
  \hline
       3.7423 \\
       2.9385 \\
       5.7125 \\
       3.4549 \\
       0.0191 \\
       1.3365 \\
       1.2082 \\
       0.9125 \\
       1.2202 \\
       \hline
       2.2406 \\
       2.0409 \\
       5.5215 \\
       3.4065 \\
       0.0008 \\
       1.4194 \\
       1.6325 \\
       0.9084 \\
       1.2028 \\
            \hline
            -2.2511\\
            1.2338\\
            \hline
            0.0182\\
\end{tabular}
&
\begin{tabular}{c}
  In-As \\
  \hline
       3.5655 \\
       2.9008 \\
       5.9270 \\
       3.4482 \\
       0.0147 \\
       1.3568 \\
       1.2211 \\
       0.8811 \\
       1.2880 \\
       \hline
       2.2916 \\
       2.1313 \\
       5.7498 \\
       3.3732 \\
       0.0023 \\
       1.3955 \\
       1.6602 \\
       0.9131 \\
       1.2172 \\
            \hline
            -2.2073\\
            1.2546\\
            \hline
            0.0096\\
\end{tabular}
&
\begin{tabular}{c}
  In-Sb \\
  \hline
       4.1432 \\
       3.7455 \\
       6.1345 \\
       4.0470 \\
       0.0017 \\
       1.3823 \\
       1.4496 \\
       0.8716 \\
       1.3863 \\
       \hline
       3.3104 \\
       3.1702 \\
       6.2579 \\
       3.8005 \\
       0.0038 \\
       1.6355 \\
       1.6168 \\
       0.9284 \\
       1.3715 \\
            \hline
            -2.9363\\
            1.2860\\
            \hline
            0.0176\\
\end{tabular}
    \\
    \hline
  \end{tabular}
  \caption{Environment dependent onsites parameters for group IV and III-V materials. In Si and Ge, both 'a' and 'c' denote the same atom.
  For Si-Ge bond, a correspond to Si and c correspond to Ge. The parameters $I$'s and $O$'s are in the unit of eV.
  parameters $\lambda$'s are in the unit of $\AA^{-1}$. The nonzero $\delta d_{ij}$ is introduced to match ETB results with experimental targets under room temperature.}\label{tab: nb to onsite, GeSi/IIIVs}
\end{table*}

\begin{table*} 
  \centering
  \begin{tabular}{c|c|c|c|c|c|c|c|c|c|c|c|c}
    \hline\hline
    \begin{tabular}{c}
    bond \\
    \hline 
      $C_{s_cp_c,a}$  \\
      $C_{p_cd_c,a}$  \\
      $C_{d_cd_c,a}$ \\
     \hline
      $C_{s_ap_a,c}$ \\
      $C_{p_ad_a,c}$ \\
      $C_{d_ad_a,c}$ \\
  \end{tabular}
  &
  \begin{tabular}{c}
    Si-Si \\
    \hline 
              1.2234 \\ 
              3.4303 \\ 
              9.9099 \\
     \hline
      \\
      \\
      \\
  \end{tabular}
  &
  \begin{tabular}{c}
    Ge-Ge \\
    \hline 
              1.1939 \\ 
              3.3684 \\ 
              9.8628 \\ 
     \hline
      \\
      \\
      \\
  \end{tabular}
  &
  \begin{tabular}{c}
    Si-Ge \\
    \hline 
              1.2030 \\
              3.3930 \\
              9.8856 \\
     \hline
              1.2030 \\
              3.3930 \\
              9.8856 \\
  \end{tabular}
  &
  \begin{tabular}{c}
    Al-P \\
    \hline 
              1.5306 \\ 
              3.5101 \\ 
              8.4800 \\ \hline
              1.2755 \\ 
              3.7066 \\ 
             10.1674 \\  
  \end{tabular}
  &
  \begin{tabular}{c}
    Ga-P \\
    \hline 
              1.2321 \\ 
              3.3655 \\ 
              8.9391 \\ \hline 
              1.2266 \\ 
              3.3529 \\ 
              9.1512 \\ 
  \end{tabular}
  &
  \begin{tabular}{c}
    In-P \\
    \hline 
              1.5843 \\ 
              3.2494 \\ 
              8.2225 \\ \hline 
              1.0321 \\ 
              3.8671 \\ 
              8.8370 \\
  \end{tabular}
  &
  \begin{tabular}{c}
    Al-As \\
    \hline 
              1.9559 \\ 
              3.6671 \\ 
              6.9304 \\ \hline
              1.4219 \\ 
              3.8677 \\ 
              9.0338 \\ 
  \end{tabular}
  &
  \begin{tabular}{c}
    Ga-As \\
    \hline 
              1.2601 \\ 
              3.4064 \\ 
              9.2562 \\ \hline
              1.2327 \\ 
              3.5647 \\ 
              9.9997 \\ 

  \end{tabular}
  &
  \begin{tabular}{c}
    In-As \\
    \hline 
              1.1396 \\ 
              3.3227 \\ 
              9.1776 \\ \hline
              1.1388 \\ 
              3.3128 \\ 
              9.7860 \\ 
  \end{tabular}
  &
  \begin{tabular}{c}
    Al-Sb \\
    \hline 
              1.5751 \\ 
              3.5628 \\ 
              8.4919 \\ \hline
              1.2914 \\ 
              3.5603 \\ 
              8.5971 \\  
  \end{tabular}
  &
  \begin{tabular}{c}
    Ga-Sb \\
    \hline 
              1.9561 \\ 
              3.8564 \\ 
              6.9425 \\ \hline
              1.9606 \\ 
              3.8573 \\ 
              6.7043 \\ 
  \end{tabular}
  &
  \begin{tabular}{c}
    In-Sb \\
    \hline 
              1.0291 \\ 
              3.3380 \\ 
              8.7305 \\ \hline
              1.1456 \\ 
              3.3593 \\ 
              8.6512 \\ 
  \end{tabular}
     \\
    \hline
  \end{tabular}
  \caption{Off-diagonal onsite parameters due to dipole and quadrupole potentials. In Si and Ge, both 'a' and 'c' denote the same atom,
  parameters $C_{\alpha_a\beta_a,c}$ are left empty due to relation $C_{\alpha_a\beta_a,c} = C_{\alpha_c\beta_c,a}$.
  For Si-Ge bond, 'a' correspond to Si and 'c' correspond to Ge. All parameters are in the unit of eV.}
  \label{tab:off_diagonal_onsite_parameters_multipole}
\end{table*}

\subsection{Multipole expansion of atomic potentials}
The local potential near atom $i$  is approximated by a summation of
the potential of atom $i$ and potential of its nearest neighbors(NNs) $j$
  \begin{equation}\label{eq:potential}
    U^{tot}_{i}\left(\mathbf{r}\right)= U_{i}\left(\left|\mathbf{r} \right|\right) + \sum_{j \in \textrm{
    NNs}}U_j\left(\left|\mathbf{r}-\mathbf{d}_{ij}\right|\right),
  \end{equation}
where the relative position between atoms $i$ and $j$ is
$\mathbf{d}_{ij}$. The potential at $\mathbf{r}$ contributed by atom at
$\mathbf{d}_{ij}$ is approximated by generalized spherical potential.
This generalized spherical potential $U_j(|\mathbf{r}-\mathbf{d}_{ij}|)$
centered at $\mathbf{d}_{ij}$  has multipole expansion given by
  \begin{equation}\label{eq:potential_expansion}
    U_j(|\mathbf{r}-\mathbf{d}_{ij}|) = \sum_{l}
    U_j^{(l)}(r,d_{ij})\sum_{m=-l}^lY_{lm}^*(\Omega_\mathbf{r})Y_{lm}(\Omega_{\mathbf{d}_{ij}}),
  \end{equation}
where $\Omega_\mathbf{r}$ and $\Omega_{\mathbf{d}_{ij}}$ stands for
angles $\theta$ and $\phi$ of vectors $\mathbf{r}$ and
$\mathbf{d}_{ij}$. The $ U^{(l)}(r,d_{ij}) $ is the radial part of
multipole potential with angular momentum $l$. By substituting
$U_j\left(\left|\mathbf{r}-\mathbf{d}_{ij}\right|\right)$ in eq
(\ref{eq:potential}) by equation (\ref{eq:potential_expansion}),
the total potential near atom $i$ given by equation (\ref{eq:potential}) can be written as a summation of multipole potentials 
\begin{equation}\label{eq:strained_potential}
     U^{tot}_i(\mathbf{r}) = \sum_{l} U_i^{(l)}(\mathbf{r}),
\end{equation}
where the multipole potentials $U_j^{(l)}(\mathbf{r})$'s are given
by
\begin{eqnarray}\label{eq:environment_dependent_potential}
   & U_i^{(0)}(\mathbf{r}) =   U_{i}\left(\left|\mathbf{r}\right|\right)+ \sum_{j}
    U_j^{(0)}(r,d_{ij})
    \nonumber  \\ \label{eq:potential,l=0}
  &  U_i^{(l)}(\mathbf{r}) =
      \sum_{m}Y_{lm}^*(\Omega_\mathbf{r}) \left( \sum_{j}  U_j^{(l)}(r,d_{ij}) Y_{lm}(\Omega_{\mathbf{d}_{ij} }
      ) \right)
      \\ \label{eq:potential,l=1} \nonumber
\end{eqnarray}
The $U_i^{(l)}$'s are summations of multipoles over nearest
neighbors. The strain induced multipole potentials up to quadrupole (with $l=2$) are considered in this work. The $U_i^{(0)}$ describes the crystal
potential under hydrostatic strain. $U_i^{(0)}$ depends only bond
lengths. For unstrained or hydrostatically strained zincblende and
diamond structures, both dipole potential $U_i^{(1)}(\mathbf{r})$
and quadrupole potential $ U_i^{(2)}(\mathbf{r})$ are zero due to the
crystal symmetry of zincblende and diamond structures. For strained
systems with traceless diagonal strain component like $\varepsilon_{xx}$,
$U_i^{(2)}(\mathbf{r})$ is induced due to angle change; while for
strained systems with off-diagonal strain component like
$\varepsilon_{xy}$, both $U_i^{(1)}(\mathbf{r})$ and $
U_i^{(2)}(\mathbf{r})$ are induced.

\subsection{Strain dependent tight binding Hamiltonian}
The strain dependent ETB Hamiltonian is constructed according to the
multipole expansion of $U^{tot}_i$. Similar to the multipole
expansion of the total potential given by eq
(\ref{eq:strained_potential}), the strain dependent ETB Hamiltonian
is written as
  \begin{equation}\label{eq:strained_Hamiltonian_multipole_expansion}
    H = H^{(0)} + H^{(1)} + H^{(2)},
  \end{equation}
where the $H^{(l)}$ depends on multipole potential $U^{(l)}(\mathbf{r})$. Matrix element
$H_{\alpha_i,\beta_j}$ is thus written as $H_{\alpha_i,\beta_j} =
H_{\alpha_i,\beta_j}^{(0)} + H_{\alpha_i,\beta_j}^{(1)} +
H_{\alpha_i,\beta_j}^{(2)}$.

\subsection{Onsite elements}
The $U_{i}^{(0)}$ has contribution from atom $i$ and its neighbors.
Similar to $U_{i}^{(0)}$, the diagonal onsite energies
$H^{(0)}_{\alpha_i,\alpha_i}$ also has contribution $E_{\alpha_i}$
from atom $i$ and contributions from its neighbors. The contribution
of neighbors to diagonal onsites energies is separated to orbital
dependent part $I_{\alpha_i,j}(d_{ij})$ and orbital independent part
$O_{i,j}(d_{ij})$. The onsite elements due to $U_{i}^{(0)}$ is given
by
\begin{equation}\label{eq:env_onsites}
H^{(0)}_{\alpha_i,\alpha_i} = E_{\alpha_i} + \sum_{j \in \textrm{NNs}}
I_{\alpha_i,j}(d_{ij}) + \sum_{j \in \textrm{NNs}}
O_{i,j}(d_{ij}),
\end{equation}
with
\begin{eqnarray}
I_{\alpha_i,j}(d_{ij}) & = & I_{\alpha_i,j} e^{-\lambda_{\alpha_i,j}( d_{ij} + \delta d_{ij}-d_0 ) } \\
O_{i,j}(d_{ij}) & = & O_{i,j} e^{-\lambda_{ij}( d_{ij}+\delta
d_{ij}-d_0)}
\end{eqnarray}
Here the $d_0$ is the reference bond length. 
In this work, the bond length of unstrained GaAs is chosen as $d_0=2.447951$.
The parameter $\delta
d_{ij}$ is introduced to modulate discrepancy between \textit{ab-initio} results and experimental results. Non-zero $\delta
d_{ij}$'s are introduced to match the ETB results in this work with experimental data under room temperature; while with zero $\delta
d_{ij}$, ETB results match the zero temperature \textit{ab-initio} results.
The term $E_{\alpha_i}$ depends on orbital and
atom type instead of material type. The summation over
$I_{\alpha_i,j}(d_{ij})$ and $O_{i,j}(d_{ij})$ are the environment
dependent part of diagonal onsite energies
$H^{(0)}_{\alpha_i,\alpha_i}$. $O_{i,j}(d_{ij})$ is used to modulate the band offset and it satisfies $O_{i,j}(d_{ij}) = O_{ji}(d_{ji})$.
Similar expression is also applied to spin-orbit coupling terms
$ \Delta^{SOC}_{i} = \Delta_{i} + \sum_{j \in \textrm{NNs}}
\Delta_{i,j}$. In this work, only spin-orbit interaction of p orbitals is considered, and the bond length dependency of $\Delta_{i,j}$ is neglected.

Due to dipole and quadrupole potentials, non-zero off-diagonal
onsite elements appear. Off-diagonal onsite elements due to
multipole potentials are given by
\begin{equation}\label{eq:off_onsites}
  E_{\alpha_i\beta_i} = \langle \psi_{\alpha_i} (\mathbf{r}) | U^{(l)}(\mathbf{r}) |\psi_{\beta_i} (\mathbf{r})
  \rangle, \quad l > 1.
\end{equation}
Since the  $U^{(l)}(\mathbf{r})$ given by eq
(\ref{eq:environment_dependent_potential}) is non-spherical, to
estimate these terms, following relation is used
\begin{equation}\label{eq:angular_gaunt}
Y_{\alpha}(\Omega)Y_{\beta}(\Omega) = \sum_{\gamma}
\mathcal{G}_{\alpha,\beta}^{\gamma} Y_{\gamma}(\Omega),
\end{equation}
where the $ \mathcal{G}_{\alpha,\beta}^{\gamma}$ is the Gaunt
coefficient \cite{Gaunt_coe} defined by
\begin{equation}\label{eq:gaunt}
\mathcal{G}_{\alpha,\beta}^{\gamma} = \int
Y_{\alpha}(\Omega)Y_{\beta}(\Omega)Y_{\gamma}^*(\Omega) d\Omega
\end{equation}
with $d\Omega = \sin{\theta}d\theta d\phi$.

With eq (\ref{eq:environment_dependent_potential}), off-diagonal
onsite elements of atom $i$ can be written as a summation of terms
depending on atom $i$ and its neighbors $j$
\begin{equation}\label{eq:off_diagonal onsite}
  E_{\alpha_i\beta_i} =\sum_{j} \mathcal{M}_{\alpha,\beta}(\mathbf{\hat{d}}_{ij})
  C^{(l)}_{\alpha_i\beta_i, j}(d_{ij}),
\end{equation}
where the $ C^{(l)}_{\alpha_i\beta_i,j}$ is the integral of radial
parts of $|\psi_{\alpha_i}\rangle$, $U^{(l)}$ and $
|\psi_{\beta_i}\rangle$, given by
  \begin{equation}\label{eq:C_onsites}
    C^{(l)}_{\alpha_i\beta_i,j} = \langle R_{\alpha_i}(r)| U^{(l)}(r,d_{ij})|R_{\beta_i}
    (r)\rangle
  \end{equation}
The $\mathcal{M}_{\alpha,\gamma}$ is given by
\begin{equation}\label{eq:D}
    \mathcal{M}^{(l)}_{\alpha, \gamma}(\mathbf{\hat{d}}_{ik}) = \sum_{m'}\mathcal{G}_{\alpha,\alpha'}^{\gamma} Y_{lm'}(\Omega_{\mathbf{d}_{ik}}), \alpha' = l,m'
\end{equation}
The explicit form of $\mathcal{M}^{(l)}_{\alpha,
\gamma}(\mathbf{\hat{d}}_{ik})$'s due to multipole potentials are
given by appendix \ref{app:D_matrix}. 

The strained onsite model by equation (\ref{eq:off_diagonal onsite}) is essentially equivalent to
the Slater Koster relations which was also used by Niquet et al \cite{Niquet_TB_model} and Boykin et al \cite{Boykin2010}. 
Onsite energies in Niquet's work depend on strains components linearly; while Boykin's onsite model uses Harrison's law. 
Differently from those previous works, the diagonal onsite energies in this work follow an exponential dependency of bond 
lengths, and the off-diagonal onsite energies depend on symmetry breaking strains linearly 
which are described by equation (\ref{eq:off_diagonal onsite}). It should be noted that, for unstrained zincblende and diamond 
structures, the $U^{(l)}=0$ for $l=1,2$ due to crystal symmetry. Consequently, the strain 
induced off-diagonal onsites $E_{\alpha_i,\beta_j}$ are all zero. The onsite energies in our model depend on the atom type 
and neighbor type instead of the material type. The atom type and bond type can be clearly defined, while the material type
can not, as demonstrated by Fig.\ref{fig_material_def}. Thus the tight binding model in this work does not have ambiguity at
the material interfaces

Since this work limits orbitals $\alpha$ and $\beta$ to s,p,d and
s*, the dipole potentials lead to non-zero off-diagonal onsite among
s-p, and p-d orbitals. While the quadrupole potential lead to
non-zero off-diagonal onsite among p-p, and d-d orbitals. Therefore, there is no confusion to use $C_{\alpha_i\beta_i,j}$ instead
of $C^{(l)}_{\alpha_i,\beta_i,j}$. Since
the strain considered in this work has amplitudes up to $4\%$, it turns out the bond length dependency of $C_{\alpha_i\beta_i,j}$ can be neglected. 
Fitting parameters for onsite elements introduced in this
work include $E_{\alpha_i}$, $I_{\alpha_i,j}$,
$\lambda_{\alpha_i,j}$ and $C_{\alpha_i\beta_i,j}$. For atoms in alloys or material interfaces, where an atom might has different type of neighbors, an averaged $C_{\alpha_i\beta_i,j}$ over neighbors $j$ is used.

\begin{table*}
  \centering
  \begin{tabular}{c|c|c|c|c|c|c|c|c|c|c|c|c}
    \hline\hline
    \begin{tabular}{c}
     bond \\
     \hline
          $V_{s_c s_a \sigma}$ \\
     $V_{s^*_c s^*_a \sigma }$ \\
       $V_{s_c s^*_a \sigma }$ \\
         $V_{s_c p_a \sigma }$ \\
       $V_{s^*_c p_a \sigma }$ \\
         $V_{s_c d_a \sigma }$ \\
       $V_{s^*_c d_a \sigma }$ \\
        $V_{p_c p_a  \sigma }$ \\
            $V_{p_c p_a \pi }$ \\
        $V_{p_c d_a \sigma  }$ \\
            $V_{p_c d_a \pi }$ \\
        $V_{d_c d_a \sigma  }$ \\
            $V_{d_c d_a \pi }$ \\
         $V_{d_c d_a \delta }$ \\
       $V_{s_a s^*_c \sigma }$ \\
         $V_{s_a p_c \sigma }$ \\
       $V_{s^*_a p_c \sigma }$ \\
         $V_{s_a d_c \sigma }$ \\
       $V_{s^*_a d_c \sigma }$ \\
        $V_{p_a d_c \sigma  }$ \\
            $V_{p_a d_c \pi }$ \\
         \hline
       $\eta_{s_c s_a \sigma}$ \\
  $\eta_{s^*_c s^*_a \sigma }$ \\
    $\eta_{s_c s^*_a \sigma }$ \\
      $\eta_{s_c p_a \sigma }$ \\
    $\eta_{s^*_c p_a \sigma }$ \\
      $\eta_{s_c d_a \sigma }$ \\
    $\eta_{s^*_c d_a \sigma }$ \\
     $\eta_{p_c p_a  \sigma }$ \\
         $\eta_{p_c p_a \pi }$ \\
     $\eta_{p_c d_a \sigma  }$ \\
         $\eta_{p_c d_a \pi }$ \\
     $\eta_{d_c d_a \sigma  }$ \\
         $\eta_{d_c d_a \pi }$ \\
      $\eta_{d_c d_a \delta }$ \\
    $\eta_{s_a s^*_c \sigma }$ \\
      $\eta_{s_a p_c \sigma }$ \\
    $\eta_{s^*_a p_c \sigma }$ \\
      $\eta_{s_a d_c \sigma }$ \\
    $\eta_{s^*_a d_c \sigma }$ \\
     $\eta_{p_a d_c \sigma  }$ \\
         $\eta_{p_a d_c \pi }$ \\
    \end{tabular}
    &
\begin{tabular}{c}
  Si-Si \\
  \hline
       -1.7377 \\
       -4.2881 \\
       -1.7587 \\
       2.9260 \\
       2.5379 \\
       -2.0901 \\
       -0.1627 \\
       3.7002 \\
       -1.2896 \\
       -0.9729 \\
       2.1919 \\
       -0.9507 \\
       1.8412 \\
       -1.3776 \\
            \\
            \\
            \\
            \\
            \\
            \\
            \\\hline
       1.5188 \\
       0.7884 \\
       0.9121 \\
       1.0267 \\
       0.6723 \\
       1.2901 \\
       0.7353 \\
       0.9903 \\
       1.3057 \\
       0.7324 \\
       0.8449 \\
       0.8837 \\
       1.4832 \\
       1.4183 \\
            \\
            \\
            \\
            \\
            \\
            \\
            \\
\end{tabular}
    &
\begin{tabular}{c}
  Ge-Ge \\
  \hline
       -1.7530 \\
       -4.4947 \\
       -1.4865 \\
       2.9146 \\
       2.3919 \\
       -1.9432 \\
       -0.1556 \\
       3.8013 \\
       -1.3517 \\
       -0.7001 \\
       2.1684 \\
       -0.4385 \\
       1.5738 \\
       -1.6745 \\
            \\
            \\
            \\
            \\
            \\
            \\
            \\
             \hline
       1.5938 \\
       0.7628 \\
       0.9936 \\
       1.1150 \\
       0.6652 \\
       1.2611 \\
       0.7792 \\
       1.0020 \\
       1.3256 \\
       0.4988 \\
       0.7391 \\
       0.6221 \\
       1.4947 \\
       1.5345 \\
            \\
            \\
            \\
            \\
            \\
            \\
            \\
\end{tabular}
    &
\begin{tabular}{c}
  Ge-Si \\
  \hline
       -1.7411 \\
       -4.6183 \\
       -1.6734 \\
       2.8349 \\
       2.5087 \\
       -2.2045 \\
       -0.2007 \\
       3.6856 \\
       -1.2686 \\
       -1.0464 \\
       1.9985 \\
       -0.3279 \\
       1.6931 \\
       -1.6394 \\
       -1.5824 \\
       2.8553 \\
       2.0593 \\
       -2.2859 \\
       -0.3354 \\
       -0.9837 \\
       2.0199 \\
 \hline
       1.5187 \\
       0.5629 \\
       1.1773 \\
       1.0444 \\
       0.7828 \\
       1.2553 \\
       0.7795 \\
       0.9412 \\
       1.2571 \\
       0.7486 \\
       0.8194 \\
       0.6172 \\
       1.4207 \\
       1.5080 \\
       0.8371 \\
       1.1317 \\
       0.9643 \\
       0.9601 \\
       0.7171 \\
       0.7872 \\
       0.8921 \\
\end{tabular}
&
\begin{tabular}{c}
  Al-P \\
  \hline
       -1.7682 \\
       -4.0139 \\
       -2.0131 \\
       2.9402 \\
       2.1206 \\
       -2.2681 \\
       -0.3042 \\
       3.5838 \\
       -1.2121 \\
       -0.7139 \\
       2.2351 \\
       -0.9666 \\
       1.9252 \\
       -1.5266 \\
       -1.2241 \\
       2.5861 \\
       2.6252  \\
       -2.1557 \\
       -0.5445 \\
       -1.2443 \\
       1.8639 \\
 \hline
       1.5395 \\
       0.7239 \\
       0.9612 \\
       1.1504 \\
       0.8908 \\
       1.0099 \\
       0.6760 \\
       0.9720 \\
       1.4131 \\
       0.7045 \\
       0.9310 \\
       0.7986 \\
       1.3402 \\
       1.3826 \\
       1.0682 \\
       1.0207 \\
       0.9204 \\
       1.1400 \\
       0.6734 \\
       0.7138 \\
       0.9125 \\
\end{tabular}
&
\begin{tabular}{c}
  Al-As \\
  \hline
       -1.8219 \\
       -4.3097 \\
       -2.0242 \\
       3.1045\\
       2.1783 \\
       -2.2634 \\
       -0.3051 \\
       3.7366 \\
       -1.3318 \\
       -0.6818 \\
       2.2795 \\
       -0.7343 \\
       1.8295 \\
       -1.6782 \\
       -1.2520 \\
       2.5919 \\
       2.6105 \\
       -2.1862 \\
       -0.4197 \\
       -1.1628 \\
       1.9673 \\
 \hline
       1.5402 \\
       0.7385 \\
       0.9635 \\
       1.1291 \\
       0.9000 \\
       0.9765 \\
       0.6901 \\
       0.9481 \\
       1.4223 \\
       0.6716 \\
       0.9336 \\
       0.8016 \\
       1.2909 \\
       1.4205 \\
       1.0682 \\
       1.0266 \\
       0.9233 \\
       1.1880 \\
       0.6640 \\
       0.7090 \\
       0.8956 \\
\end{tabular}
&
\begin{tabular}{c}
  Al-Sb \\
  \hline
       -2.1063 \\
       -4.2962 \\
       -1.8153 \\
       3.3534 \\
       2.2283 \\
       -2.4048 \\
       -0.3387 \\
       4.1011 \\
       -1.6433 \\
       -0.9318 \\
       2.4007 \\
       -0.7374 \\
       1.7864 \\
       -1.8053 \\
       -1.5371 \\
       2.9884 \\
       2.5435 \\
       -2.0941 \\
       -0.2418 \\
       -0.9421 \\
       2.0986 \\
 \hline
       1.5484 \\
       0.6720 \\
       1.0249 \\
       0.9883 \\
       0.9711 \\
       0.8921 \\
       0.6394 \\
       0.9539 \\
       1.3508 \\
       0.5149 \\
       0.9104 \\
       0.8906 \\
       1.2642 \\
       1.5074 \\
       1.0043 \\
       1.0507 \\
       0.8024 \\
       1.2410 \\
       0.6954 \\
       0.7175 \\
       0.7612 \\
\end{tabular}
&
\begin{tabular}{c}
  Ga-P \\
  \hline
       -1.7010 \\
       -4.1464 \\
       -1.8778 \\
       2.8997 \\
       2.0854 \\
       -2.2303 \\
       -0.2808 \\
       3.5451 \\
       -1.1631 \\
       -0.8561 \\
       2.1997 \\
       -0.4721 \\
       1.5643 \\
       -1.4702 \\
       -1.1986 \\
       2.6045 \\
       2.6205 \\
       -1.7346 \\
       -0.4906 \\
       -0.7510 \\
       1.8737 \\
 \hline
       1.5399 \\
       0.7270 \\
       0.9639 \\
       1.0862 \\
       0.8632 \\
       1.1882 \\
       0.6625 \\
       0.9887 \\
       1.4554 \\
       0.6995 \\
       0.9056 \\
       0.7629 \\
       1.4121 \\
       1.4383 \\
       0.9752 \\
       1.0821 \\
       0.9074 \\
       1.1570 \\
       0.6609 \\
       0.7059 \\
       0.9149 \\
\end{tabular}
&
\begin{tabular}{c}
  Ga-As \\
  \hline
       -1.7842 \\
       -4.3164 \\
       -1.8820 \\
       2.9935\\
       2.1256 \\
       -2.1456 \\
       -0.2812 \\
       3.7312 \\
       -1.2992 \\
       -0.7416 \\
       2.2874 \\
       -0.4906 \\
       1.4887 \\
       -1.6107 \\
       -1.1588 \\
       2.7008 \\
       2.5674 \\
       -1.9422 \\
       -0.3828 \\
       -0.6656 \\
       2.0486 \\
\hline
       1.5565 \\
       0.7447 \\
       0.9515 \\
       1.1004 \\
       0.7836 \\
       1.1300 \\
       0.6818 \\
       0.9646 \\
       1.3846 \\
       0.6976 \\
       0.8730 \\
       0.6990 \\
       1.2959 \\
       1.4491 \\
       0.9898 \\
       1.1126 \\
       0.8269 \\
       1.0945 \\
       0.6838 \\
       0.6976 \\
       0.8941 \\
\end{tabular}
&
\begin{tabular}{c}
  Ga-Sb \\
  \hline
       -2.0232 \\
       -4.2066 \\
       -1.7410 \\
       3.2439 \\
       2.4986 \\
       -2.2758 \\
       -0.1848 \\
       4.1685 \\
       -1.5846 \\
       -1.1356 \\
       2.3716 \\
       -0.5153 \\
       1.6402 \\
       -1.8241 \\
       -1.6281 \\
       3.0092 \\
       2.2691 \\
       -2.1687 \\
       -0.3829 \\
       -0.3859 \\
       2.1917 \\
        \hline
       1.5076 \\
       0.6439 \\
       1.0117 \\
       1.0413 \\
       0.9136 \\
       1.1453 \\
       0.6042 \\
       1.0211 \\
       1.4392 \\
       0.5096 \\
       0.9348 \\
       0.6763 \\
       1.4977 \\
       1.4208 \\
       0.9824 \\
       1.0806 \\
       0.8240 \\
       0.9333 \\
       0.7762 \\
       0.7726 \\
       0.8046 \\
\end{tabular}
&
\begin{tabular}{c}
  In-P \\
  \hline
       -1.9110 \\
       -3.7944 \\
       -2.2047 \\
       3.0736 \\
       2.2361 \\
       -2.2543 \\
       -0.3446 \\
       3.6073 \\
       -1.2755 \\
       -0.5488 \\
       2.2517 \\
       -0.4615 \\
       1.6186 \\
       -1.6310 \\
       -1.1401 \\
       2.5465 \\
       2.6249\\
       -1.6800 \\
       -0.7584 \\
       -0.5816 \\
       1.8626 \\
        \hline
       1.5274 \\
       0.7325 \\
       0.9559 \\
       1.0960 \\
       0.8578 \\
       1.1067 \\
       0.6949 \\
       1.0454 \\
       1.4932 \\
       0.7044 \\
       0.8241 \\
       0.8025 \\
       1.3955 \\
       1.3471 \\
       0.9630 \\
       1.0298 \\
       0.8790 \\
       1.0923 \\
       0.6906 \\
       0.7041 \\
       0.9100 \\
\end{tabular}
&
\begin{tabular}{c}
  In-As \\
  \hline
       -1.9667 \\
       -4.2049 \\
       -2.1482 \\
       3.2715\\
       2.2493 \\
       -2.2986 \\
       -0.2867 \\
       3.9261 \\
       -1.4074 \\
       -0.6025 \\
       2.2879 \\
       -0.4708 \\
       1.6103 \\
       -1.8837 \\
       -1.1581 \\
       2.6184\\
       2.6070 \\
       -1.7252 \\
       -0.4789 \\
       -0.5791 \\
       1.9421 \\
            \hline
       1.5436 \\
       0.7794 \\
       0.9384 \\
       1.0707 \\
       0.8618 \\
       1.0693 \\
       0.6982 \\
       1.0434 \\
       1.4411 \\
       0.6964 \\
       0.7977 \\
       0.8020 \\
       1.4221 \\
       1.3581 \\
       0.9941 \\
       1.0809 \\
       0.8193 \\
       1.1253 \\
       0.6837 \\
       0.6993 \\
       0.9198 \\
\end{tabular}
&
\begin{tabular}{c}
  In-Sb \\
  \hline
       -2.2797 \\
       -4.1696 \\
       -1.8748 \\
       3.5395\\
       2.2701\\
       -2.4392 \\
       -0.1813 \\
       4.2661\\
       -1.7708 \\
       -0.9446 \\
       2.4045\\
       -0.6675 \\
       1.7524 \\
       -2.0733 \\
       -1.3964 \\
       3.0903 \\
       2.3266 \\
       -2.0149 \\
       -0.3659 \\
       -0.3351 \\
       2.0716 \\
\hline
       1.5461 \\
       0.6794 \\
       0.9793 \\
       1.0835 \\
       0.9525 \\
       0.9973 \\
       0.7439 \\
       0.9518 \\
       1.4457 \\
       0.5439 \\
       0.8398 \\
       0.7115 \\
       1.3794 \\
       1.2748 \\
       0.9732 \\
       1.1634 \\
       0.7068 \\
       0.9660 \\
       0.7474 \\
       0.7927 \\
       0.8251 \\
\end{tabular}
    \\
\hline
  \end{tabular}
  \caption{Bond length dependent interactomic coupling parameters for group IV and III-V materials.
  In Si and Ge,  both 'a' and 'c' denote the same atom.
  For Si-Ge bond, 'a' correspond to Si and 'c' correspond to Ge. The parameters $V$'s are in the unit of eV.
  parameters $\eta$'s are in the unit of $\AA^{-1}$. }\label{tab:interatomic_parameters_bond_length_SiGe IIIVs}
\end{table*}

\begin{table*} 
  \centering
  \begin{tabular}{c|c|c|c|c|c|c|c|c|c|c|c|c}
    \hline\hline
    \begin{tabular}{c}
    bond\\
    \hline
     $P_{s_a p_c \sigma }$      \\ 
     $P_{s_a d_c \sigma }$      \\ 
     $P_{p_a p_c  \sigma }$     \\
     $P_{p_a p_c \pi }$         \\ 
     $P_{s_c p_a \sigma }$      \\ 
     $P_{s_c d_a \sigma }$      \\ 
    \hline
     $S_{s_a p_c \sigma }$      \\ 
     $S_{s_a d_c \sigma }$      \\ 
     $S_{p_a p_c  \sigma }$     \\
     $S_{p_a p_c \pi }$         \\ 
     $S_{d_a d_c \sigma  }$     \\
     $S_{d_a d_c \pi }$         \\
     $S_{d_a d_c \delta }$      \\
     $S_{s_c p_a \sigma }$      \\ 
     $S_{s_c d_a \sigma }$      \\ 
    \hline
     $Q_{s_a p_c \sigma }$      \\ 
     $Q_{s_a d_c \sigma }$      \\ 
     $Q_{p_a p_c  \sigma }$     \\
     $Q_{p_a p_c \pi }$         \\ 
     $Q_{d_a d_c \sigma  }$     \\
     $Q_{d_a d_c \pi }$         \\
     $Q_{d_a d_c \delta }$      \\
     $Q_{s_c p_a \sigma }$      \\ 
     $Q_{s_c d_a \sigma }$      \\ 
  \end{tabular}
   &
    \begin{tabular}{c}
    Si-Si \\
    \hline
             -1.5396 \\ 
              0.7752 \\ 
             -0.9283 \\ 
              1.6156 \\ 
              \\
              \\
    \hline
              0.7491 \\ 
              1.4609 \\ 
              1.6103 \\ 
             -3.8712 \\ 
              0.7450 \\ 
              4.0875 \\ 
              3.9344 \\ 
              \\
              \\
    \hline
              6.5771 \\ 
             -1.3985 \\ 
             -2.5641 \\ 
             -0.9290 \\ 
              1.9700 \\ 
              6.9775 \\ 
             -0.4367 \\ 
             \\
             \\
  \end{tabular}
   &
  \begin{tabular}{c}
    Ge-Ge \\
    \hline
             -1.5663 \\ 
              0.7925 \\ 
             -0.6865 \\ 
              1.2451 \\              
              \\
              \\
    \hline
              0.8861 \\ 
              1.5098 \\ 
              1.6759 \\ 
             -2.6283 \\ 
              0.6304 \\ 
              3.2465 \\ 
              3.2883 \\                
              \\
              \\
    \hline    
              5.1614 \\ 
             -1.4161 \\ 
             -1.9725 \\ 
             -0.7786 \\ 
              2.0320 \\ 
              6.8269 \\ 
             -0.4345 \\ 
             \\
             \\
  \end{tabular}
   &
  \begin{tabular}{c}
    Si-Ge \\
    \hline
             -1.5006 \\
              0.8145 \\
             -0.7794 \\
              1.5188 \\
             -1.5006 \\
              0.8145 \\
    \hline
              0.8200 \\
              1.4848 \\
              1.4812 \\
             -3.4877 \\
              0.7508 \\
              3.8909 \\
              3.7768 \\
              0.8200 \\
              1.4848 \\
    \hline
              6.2119 \\
             -1.3773 \\
             -2.2944 \\
             -0.9155 \\
              2.0051 \\
              6.9180 \\
             -0.2475 \\
              6.2119 \\
             -1.3773 \\
  \end{tabular}
   &
   \begin{tabular}{c}
    Al-P \\
    \hline             
             -1.6592 \\ 
              0.3091 \\ 
             -1.0469 \\ 
              1.8003 \\ 
             -2.3325 \\ 
              0.3045 \\
    \hline              
              1.9743 \\ 
              1.5210 \\ 
              3.1829 \\ 
             -4.5544 \\ 
              0.9623 \\ 
              3.6546 \\ 
              3.7809 \\ 
              1.1003 \\ 
              2.3270 \\
    \hline              
              6.5773 \\ 
             -2.2243 \\ 
             -2.6508 \\ 
             -0.4430 \\ 
              2.0628 \\ 
              6.3774 \\ 
             -0.8822 \\ 
              7.3014 \\ 
             -1.6439 \\ 
  \end{tabular}
   &
       \begin{tabular}{c}
    Ga-P \\
    \hline             
             -1.5167 \\ 
              0.7372 \\ 
             -0.3635 \\ 
              1.6262 \\ 
             -1.5468 \\ 
              0.6986 \\ 
    \hline             
              0.7668 \\ 
              1.4103 \\ 
              2.0255 \\ 
             -4.4913 \\ 
              0.7014 \\ 
              3.7256 \\ 
              4.0881 \\ 
              0.7325 \\ 
              1.4101 \\ 
    \hline              
              5.6126 \\ 
             -1.0040 \\ 
             -2.3040 \\ 
             -0.5811 \\ 
              2.0977 \\ 
              6.1846 \\ 
             -0.2823 \\ 
              6.3718 \\ 
             -1.3167 \\ 
  \end{tabular}
   &
       \begin{tabular}{c}
    In-P \\
    \hline             
             -1.3417 \\ 
              1.1406 \\ 
             -0.2978 \\ 
              1.0269 \\ 
             -2.7879 \\ 
              0.4468 \\ 
    \hline              
              1.7468 \\ 
              1.3737 \\ 
              2.6098 \\ 
             -4.4096 \\ 
              0.2811 \\ 
              3.7468 \\ 
              3.5400 \\ 
              0.8607 \\ 
              2.0861 \\ 
    \hline              
              4.3389 \\ 
             -2.1996 \\ 
             -2.3389 \\ 
             -0.0174 \\ 
              2.3898 \\ 
              7.1001 \\ 
             -0.7120 \\ 
              6.0554 \\ 
             -1.3153 \\ 
  \end{tabular}
   &
   \begin{tabular}{c}
    Al-As \\
    \hline             
             -0.9448 \\ 
              1.0546 \\ 
             -1.5842 \\ 
              1.3440 \\ 
             -2.6736 \\ 
              0.3458 \\ 
    \hline              
              1.8554 \\ 
              1.8819 \\ 
              2.8324 \\ 
             -4.3925 \\ 
              0.7210 \\ 
              4.2782 \\ 
              3.7232 \\ 
              0.4794 \\ 
              2.3560 \\
    \hline
              4.6049 \\ 
             -2.0707 \\ 
             -2.6020 \\ 
             -0.0880 \\ 
              2.4063 \\ 
              5.9092 \\ 
             -1.3089 \\ 
              7.0724 \\ 
             -0.8685 \\ 
  \end{tabular}
   &
  \begin{tabular}{c}
    Ga-As \\
    \hline             
             -1.3555 \\ 
              0.5127 \\ 
             -0.4684 \\ 
              1.0823 \\ 
             -2.2816 \\ 
              0.5314 \\
    \hline              
              1.4927 \\ 
              1.8221 \\ 
              1.8866 \\ 
             -4.2555 \\ 
              0.7340 \\ 
              3.1996 \\ 
              3.6569 \\ 
              0.5577 \\ 
              2.2435 \\ 
    \hline              
              5.2229 \\ 
             -1.5659 \\ 
             -1.2315 \\ 
             -1.1158 \\ 
              2.4369 \\ 
              7.0035 \\ 
             -0.7043 \\ 
              6.1072 \\ 
             -1.0584 \\ 
  \end{tabular}
   &
       \begin{tabular}{c}
    In-As \\
    \hline             
             -1.5821 \\ 
              1.0312 \\ 
             -0.4048 \\ 
              0.7503 \\ 
             -1.7354 \\ 
              0.7030 \\ 
    \hline              
              1.2099 \\ 
              1.5578 \\ 
              2.4993 \\ 
             -4.2825 \\ 
              0.6007 \\ 
              3.4492 \\ 
              3.9674 \\ 
              0.7424 \\ 
              1.4634 \\ 
    \hline              
              4.7711 \\ 
             -1.4199 \\ 
             -1.1147 \\ 
             -0.7130 \\ 
              1.7380 \\ 
              6.9446 \\ 
             -0.4713 \\ 
              5.4079 \\ 
             -1.1532 \\ 
  \end{tabular}
   &     \begin{tabular}{c}
   Al-Sb \\
    \hline             
             -1.1122 \\ 
              0.8430 \\ 
             -1.1008 \\ 
              0.4365 \\ 
             -2.4051 \\ 
              0.4151 \\ 
    \hline              
              1.5753 \\ 
              1.4029 \\ 
              2.7592 \\ 
             -3.6569 \\ 
              0.4945 \\ 
              2.9925 \\ 
              2.8669 \\ 
              0.7568 \\ 
              2.2419 \\
    \hline              
              4.7149 \\ 
             -1.5381 \\ 
             -2.4559 \\ 
              0.1086 \\ 
              1.9471 \\ 
              6.3621 \\ 
             -1.2464 \\ 
              5.3544 \\ 
             -0.7379 \\ 
  \end{tabular}
   &
       \begin{tabular}{c}
    Ga-Sb \\
    \hline
             -1.2324 \\ 
              0.6116 \\ 
             -0.9431 \\ 
              0.4087 \\ 
             -2.6815 \\ 
              0.4081 \\  
    \hline              
              1.1479 \\ 
              1.8530 \\ 
              2.2525 \\ 
             -3.4164 \\ 
              0.1092 \\ 
              4.0625 \\ 
              2.6015 \\ 
              0.1379 \\ 
              2.3834 \\ 
    \hline              
              3.6729 \\ 
             -2.0985 \\ 
             -1.3465 \\ 
             -0.6194 \\ 
              2.3476 \\ 
              5.9208 \\ 
             -1.1962 \\ 
              6.9797 \\ 
             -0.7086 \\ 
  \end{tabular}
  &
       \begin{tabular}{c}
    In-Sb \\
    \hline             
             -1.0398 \\ 
              1.0719 \\ 
             -0.6557 \\ 
              0.5719 \\ 
             -1.8459 \\ 
              0.8668 \\ 
    \hline              
              1.9147 \\ 
              0.7074 \\ 
              2.7781 \\ 
             -2.8257 \\ 
              0.3453 \\ 
              3.3261 \\ 
              3.6276 \\ 
             -0.1957 \\ 
              2.0629 \\
    \hline              
              3.5491 \\ 
             -1.4815 \\ 
             -1.8447 \\ 
             -0.0713 \\ 
              2.2434 \\ 
              6.6177 \\ 
             -0.6252 \\ 
              5.2978 \\ 
             -1.1249 \\
  \end{tabular}
   \\
    \hline
  \end{tabular}
  \caption{Interatomic coupling due to dipole and quadrupole potentials. In Si and Ge, both 'a' and 'c' denote the same atom.
  For Si-Ge bond, a correspond to Si and c correspond to Ge. All parameters are in the unit of eV.}
  \label{tab:interatomic_coupling_parameters_multipole}
\end{table*}

\subsection{Interatomic couplings}
Interatomic couplings $H^{(0)}_{\alpha_i,\beta_j}$ due to $U^{(0)}$
which couple orbital $\alpha$ of atom $i$ and orbital $\beta$ of
atom $j$ follows the Slater Koster formulas
\cite{Slater_Tightbinding,Podolskiy_TBElements}. Bond length
dependent two center integrals in this work are approximated by
exponential law
\begin{equation}\label{eq:coupling}
V_{\alpha_i \beta_j |m|}(d_{ij}) =  V_{\alpha_i \beta_j|m|}
e^{-\eta_{\alpha_i \beta_j |m|}(d_{ij}+\delta d_{ij}-d_0)}.
\end{equation}
The $\delta d_{ij}$ is the parameter introduced in order to match
the ETB band structure with experimental results.

The interatomic coupling due to multipole potential $U^{(l)}$ are
written as
\begin{equation}\label{eq:environment_dependent_H}
V^{(l)}_{\alpha_i,\beta_j} = \langle \psi_{\alpha}(\mathbf{r}) |
U^{(l)}(\mathbf{r}) + U^{(l)}(\mathbf{r}-\mathbf{d}_{ij} ) |
\psi_{\beta} (\mathbf{r}- \mathbf{d}_{ij}) \rangle.
\end{equation}
By substituting $U^{(l)}$ with equation
(\ref{eq:environment_dependent_potential}), this integral can be
written as
\begin{eqnarray}\label{eq:interatomic_coupling_multipole}
  V^{(l)}_{\alpha_i,\beta_j} = &  \sum_{\gamma,k} \mathcal{M}^{(l)}_{\alpha,\gamma}(\mathbf{\hat{d}}_{ik}) Q^{(l)}_{\gamma_i,\beta_j}(d_{ik}) +  \\
     & \sum_{\gamma',k'} Q^{(l)}_{\alpha_i,\gamma'_j}(d_{jk'}) \mathcal{M}^{(l)}_{\gamma',\beta}(\mathbf{\hat{d}}_{jk'})
 \nonumber
\end{eqnarray}
where the $k$ denotes the nearest neighbors of atom $i$ and the $k'$ denotes
the nearest neighbors of atom $j$. The
$Q^{(l)}_{\gamma,\beta}(d_{ik})$ and
$Q^{(l)}_{\alpha,\gamma'}(d_{jk'})$ are given by
\begin{eqnarray}\label{eq:interatomic_coupling_Q}
Q^{(l)}_{\gamma_i,\beta_j}(d_{ik}) & = &  \langle \psi_{\gamma}(\mathbf{r}) | U^{(l)}(r,d_{ik}) | \psi_{\beta}(\mathbf{r}-\mathbf{d}_{ij}) \rangle \\
Q^{(l)}_{\alpha_i,\gamma'_j}(d_{jk'}) & = & \langle
\psi_{\alpha}(\mathbf{r}) |
U^{(l)}(|\mathbf{r}-\mathbf{d}_{ij}|,d_{jk'}) |
\psi_{\gamma'}(\mathbf{r}-\mathbf{d}_{ij}) \rangle \nonumber
\end{eqnarray}
The $| \psi_{\gamma}(\mathbf{r}) \rangle$ has the same radial part
as $| \psi_{\alpha}(\mathbf{r}) \rangle$ , although $\gamma$ and
$\alpha$ are different.
$Q^{(l)}_{\gamma_i,\beta_j}(d_{ik})$ and
$Q^{(l)}_{\alpha_i,\gamma'_j}(d_{jk'})$ are three center integrals
involving orbitals of atom $i$,$j$ and potential $U^{(l)}$ from atom
$k$ or $k'$. However, since the quadrupole potential $U^{(l)}$ are
centered either at atom $i$ or $j$, the
$Q^{(l)}_{\gamma_i,\beta_j}(d_{ik})$ and
$Q^{(l)}_{\alpha_i,\gamma'_j}(d_{jk'})$ has the expression of two
center integrals describing by Slater Koster formulas. To simplify
the formula, we approximate the  effect of $U^{(l)}(r,d_{ik})$'s by
using averaged potential over $k$ and $k'$ to remove the dependency
of atom $k$ and $k'$, $\bar{U}^{(l)}(r) = \frac{1}{n_{k}}\sum_{k}
U^{(l)}(r,d_{ik})$ , $\bar{U}^{(l)}(|\mathbf{r}-\mathbf{d}_{ij}|) =
\frac{1}{n_{k'}}\sum_{k'}
U^{(l)}(|\mathbf{r}-\mathbf{d}_{ij}|,d_{jk'}) $.
Similar to the onsite energies, the strain 
induced terms $V^{(l)}_{\alpha_i,\beta_j}$ are all zero for unstrained bulk zincblende 
and diamond materials.

For dipole potentials, the complete explicit expression of
equation (\ref{eq:interatomic_coupling_multipole}) is lengthy.
In this work, we find it is sufficient to approximated equation (\ref{eq:interatomic_coupling_multipole}) with
Slater Koster formula for dipole potentials. 
The $U^{(1)}$ introduces
strain correction $\delta V^{(1)}_{\alpha_i \beta_j |m|}$ to interatomic interaction parameters $V_{\alpha_i \beta_j |m|}(d_{ij})$ given by equation
(\ref{eq:coupling}). The $\delta V^{(1)}_{\alpha_i \beta_j |m|}$  
has the expression
\begin{equation}\label{eq:coupling_dipole}
    \delta V^{(1)}_{\alpha_i \beta_j |m|} =
    \frac{4\pi}{3}P_{\alpha_i,\beta_j,|m|} \left( p_{ij} +p_{ji }\right) + \frac{4\pi}{3}S_{\alpha_i,\beta_j,|m|} \left(
    q_{ij} +q_{ji}\right),
\end{equation}
where the $p_{ij}$ and $q_{ij}$ estimate the dipole potential along
bond $\mathbf{d}_{ij}$. $P_{\alpha_i,\beta_j,|m|}$ and
$S_{\alpha_i,\beta_j,|m|}$ are fitting parameters. $p_{ij}$ and
$q_{ij}$ are given as
\begin{eqnarray}\label{eq:p_ij}
    p_{ij} & = &
\sum_{k,m} Y_{1,m}\left(
\Omega_{\mathbf{d}_{i,k}}\right)Y_{1,m}\left(
\Omega_{\mathbf{d}_{i,j}}\right)
    \\
    q_{ij} & = &
    \sum_{k,m} Y_{1,m}\left( \Omega_{\mathbf{d}_{i,k}}\right)Y_{1,m}\left( \Omega_{\mathbf{d}_{i,j}}\right)
    \frac{ \delta \mathbf{d}_{ik} }{\bar{d}}.\\  \nonumber
\end{eqnarray}
\begin{eqnarray}\label{eq:p_ji}
     p_{ji} & = &
\sum_{k',m} Y_{1,m}\left(
\Omega_{\mathbf{d}_{j,k}}\right)Y_{1,m}\left(
\Omega_{\mathbf{d}_{j,i}}\right)
    \\
    q_{ji} & = &
    \sum_{k',m} Y_{1,m}\left( \Omega_{\mathbf{d}_{j,k'}}\right)Y_{1,m}\left( \Omega_{\mathbf{d}_{j,i}}\right)
    \frac{ \delta \mathbf{d}_{jk'} }{\bar{d}}.  \nonumber
\end{eqnarray}
The $\bar{d}$ is the average bond length. More discussion of his
approximation is given in appendix
\ref{app:prove_approximation_dipole}. $p_{ij}$ and $q_{ij}$ estimate
the impact of dipole moment to neighbors. The non-zero $p_{ij}$
correspond to non-zero off-diagonal strain components, while the
nonzero term $q_{ij}$ corresponds to bond length changes which break
crystal symmetry.

For quadrupole potentials, we find it is sufficient to drop the bond
length dependency of $\bar{U}^{(2)}(r)$ and
$\bar{U}^{(l)}(|\mathbf{r}-\mathbf{d}_{ij}|)$ from equation
(\ref{eq:interatomic_coupling_Q}) since we consider strain up to
$4\%$ in this work. Thus $Q_{\gamma_i,\beta_j}(d_{ik}) $ and
$Q_{\alpha_i,\gamma'_j}(d_{k'j})$  can be simplified by
\begin{eqnarray}
Q_{\gamma_i,\beta_j} & = &  \langle \psi_{\gamma}(\mathbf{r}) | \bar{U}^{(2)}(r)  | \psi_{\beta}(\mathbf{r}-\mathbf{d}_{ij}) \rangle \\
Q_{\alpha_i,\gamma'_j} & = & \langle \psi_{\alpha}(\mathbf{r}) |
\bar{U}^{(2)}(|\mathbf{r}-\mathbf{d}_{ij}|) |
\psi_{\gamma'}(\mathbf{r}-\mathbf{d}_{ij}) \rangle
\end{eqnarray}
Here the fitting parameters in Slater Koster form $ Q_{\alpha_i,\beta_j,
|m| } $ are introduced.

\begin{figure}
\includegraphics[width=0.4\textwidth]{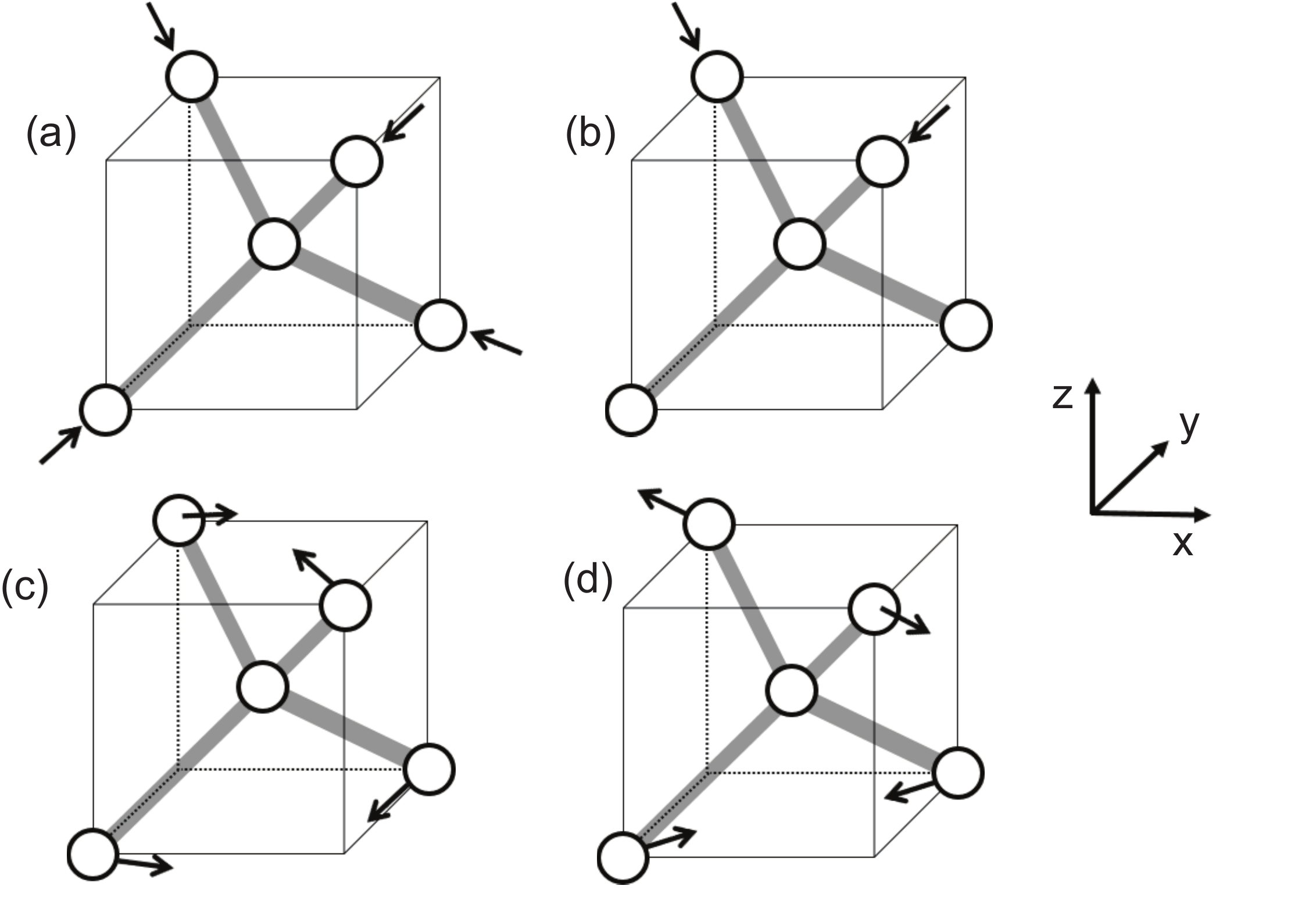}
\caption{Strained systems considered in this work. (a) hydrostatic strain, (b) with two bond length changes,
(c) diagonal strain with $\varepsilon_{xx} = \varepsilon_{yy} = -0.5 \varepsilon_{zz}$, (d) off-diagonal strain with $\varepsilon_{xy} \neq 0$}\label{fig_strain_displacement}
\end{figure}

\begin{figure}
\includegraphics[width=0.5\textwidth]{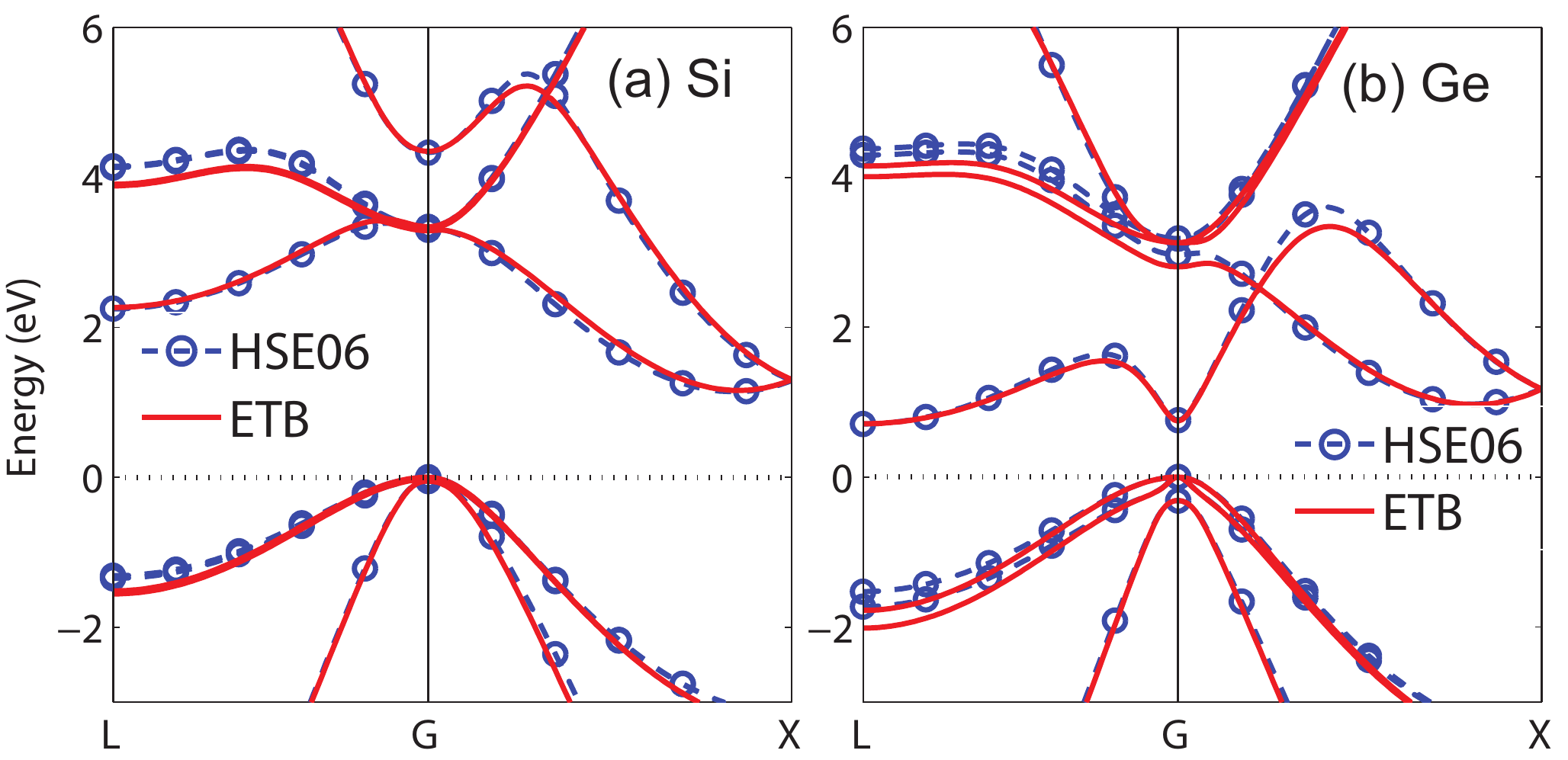}
\caption{Band structure of III-Vs materials with ETB and HSE06
calculations. Presented band structures of IV materials include Si (a) and Ge (b). ETB band structures are in good agreement with HSE06 results. The HSE06 bands are adjusted to match experimental results under room temperature.  }\label{fig_IVs}
\end{figure}

\begin{table}[h]
    \small
    \setlength{\tabcolsep}{1pt}
\begin{tabular}
[c]{c|c|c}\hline\hline
& Si & Ge  \\ \hline%
\begin{tabular}
[c]{c}%
targets\\\hline
$E_{g}(\Gamma)$\\
$E_{g}(X)$\\
$E_{g}(L)$\\
$\Delta_{SO}$\\
\\
$m_{hh100}$\\
$m_{hh110}$\\
$m_{hh111}$\\
$m_{lh100}$\\
$m_{lh110}$\\
$m_{lh111}$\\
$m_{so}$\\%
\\
$m_{c\Gamma}$\\%
$m_{cXl}$\\
$m_{cXt}$\\
$m_{cLl}$\\
$m_{cLt}$\\
\end{tabular}
&
\begin{tabular}
[c]{cc}%
\begin{tabular}
[c]{ccc}%
HSE06 & ETB & error\\\hline
          $3.301$ &           $3.332$ &                $0.9\%$ \\
          $1.141$ &           $1.155$ &                $1.2\%$ \\
          $2.246$ &           $2.245$ &                $0.1\%$ \\
          $0.051$ &           $0.051$ &                $0.0\%$ \\
  &  & \\
%
             $0.260$ &              $0.266$ &              $2.5\%$ \\ 
             $0.522$ &              $0.535$ &              $2.4\%$ \\ 
             $0.649$ &              $0.672$ &              $3.5\%$ \\ 
             $0.190$ &              $0.179$ &              $5.9\%$ \\ 
             $0.139$ &              $0.134$ &              $3.7\%$ \\ 
             $0.132$ &              $0.127$ &              $3.6\%$ \\ 
             $0.225$ &              $0.218$ &              $2.8\%$ \\ 
             & & \\               
             &              - &              - \\%
             $0.856$ &              $0.754$ &             $11.9\%$ \\ 
             $0.191$ &              $0.194$ &              $1.2\%$ \\ 
             $1.641$ &              $1.774$ &              $8.1\%$ \\ 
             $0.130$ &              $0.147$ &             $13.2\%$ \\ 
\end{tabular}
&
\begin{tabular}
[c]{c}%
Ref \\
\hline
                        $3.34$ \\ 
                        $1.12$ \\ 
                        $2.04$ \\ 
                        $0.04$ \\ 
 \\ 
                        $0.29$ \\ 
                        $0.54$ \\ 
                        $0.75$ \\ 
                        $0.20$ \\ 
                        $0.15$ \\ 
                        $0.14$ \\ 
                        $0.23$ \\ 
                        \\
                       $-$ \\ 
                        $0.91$ \\ 
                        $0.19$ \\ 
                        $3.43$ \\ 
                        $0.17$ \\ 
\end{tabular}
\\
\end{tabular}
&
\begin{tabular}
[c]{cc}%
\begin{tabular}
[c]{ccc}%
HSE06 & ETB & error \\\hline
          $0.755$ &           $0.744$ &                $1.4\%$ \\
          $0.974$ &           $0.945$ &                $3.0\%$ \\
          $0.709$ &           $0.678$ &                $4.4\%$ \\
          $0.313$ &           $0.311$ &                $0.4\%$ \\
  &  & \\
             $0.203$ &              $0.197$ &              $2.7\%$ \\ 
             $0.378$ &              $0.381$ &              $0.6\%$ \\ 
             $0.506$ &              $0.523$ &              $3.2\%$ \\ 
             $0.040$ &              $0.040$ &              $1.0\%$ \\ 
             $0.037$ &              $0.037$ &              $0.3\%$ \\ 
             $0.035$ &              $0.035$ &              $0.2\%$ \\ 
             $0.093$ &              $0.091$ &              $2.2\%$ \\ 
			& & \\
             $0.032$ &              $0.033$ &              $3.7\%$ \\ 
             $0.840$ &              $0.768$ &              $8.5\%$ \\ 
             $0.189$ &              $0.203$ &              $7.5\%$ \\ 
             $1.577$ &              $1.738$ &             $10.2\%$ \\ 
             $0.081$ &              $0.101$ &             $23.8\%$ \\
\end{tabular}
&
\begin{tabular}
[c]{c}%
Ref \\
\hline
                        $0.81$ \\ 
                        $0.90$ \\ 
                        $0.66$ \\ 
                        $0.30$ \\ 
 \\ 
                        $0.21$ \\ 
                        $0.37$ \\ 
                        $0.51$ \\ 
                        $0.05$ \\ 
                        $0.04$ \\ 
                        $0.04$ \\ 
                        $0.10$ \\
                       \\ 
                       $-$ \\ 
                        $0.90$ \\ 
                        $0.20$ \\ 
                        $1.59$ \\ 
                        $0.08$ \\
\end{tabular}
\\
\end{tabular}
\\\hline
\end{tabular}
\centering \caption{Targets comparison of bulk XP. Critical band
edges and effective masses at $\Gamma$, $X$ and $L$ from ETB and
HSE06 calculations are compared. The $E_{g}$ and $\Delta_{SO}$ are
in the unit of eV; effective masses are scaled by free electron mass
$m_0$. The error column summarizes the relative discrepancies between HSE06
and ETB results. The Reference bandedge and effective masses are from Ref \onlinecite{IV_mass_para}.
}%
\label{tab:targets_comparison_SiGe} %
\end{table}

\begin{figure*}
\includegraphics[width=0.7\textwidth]{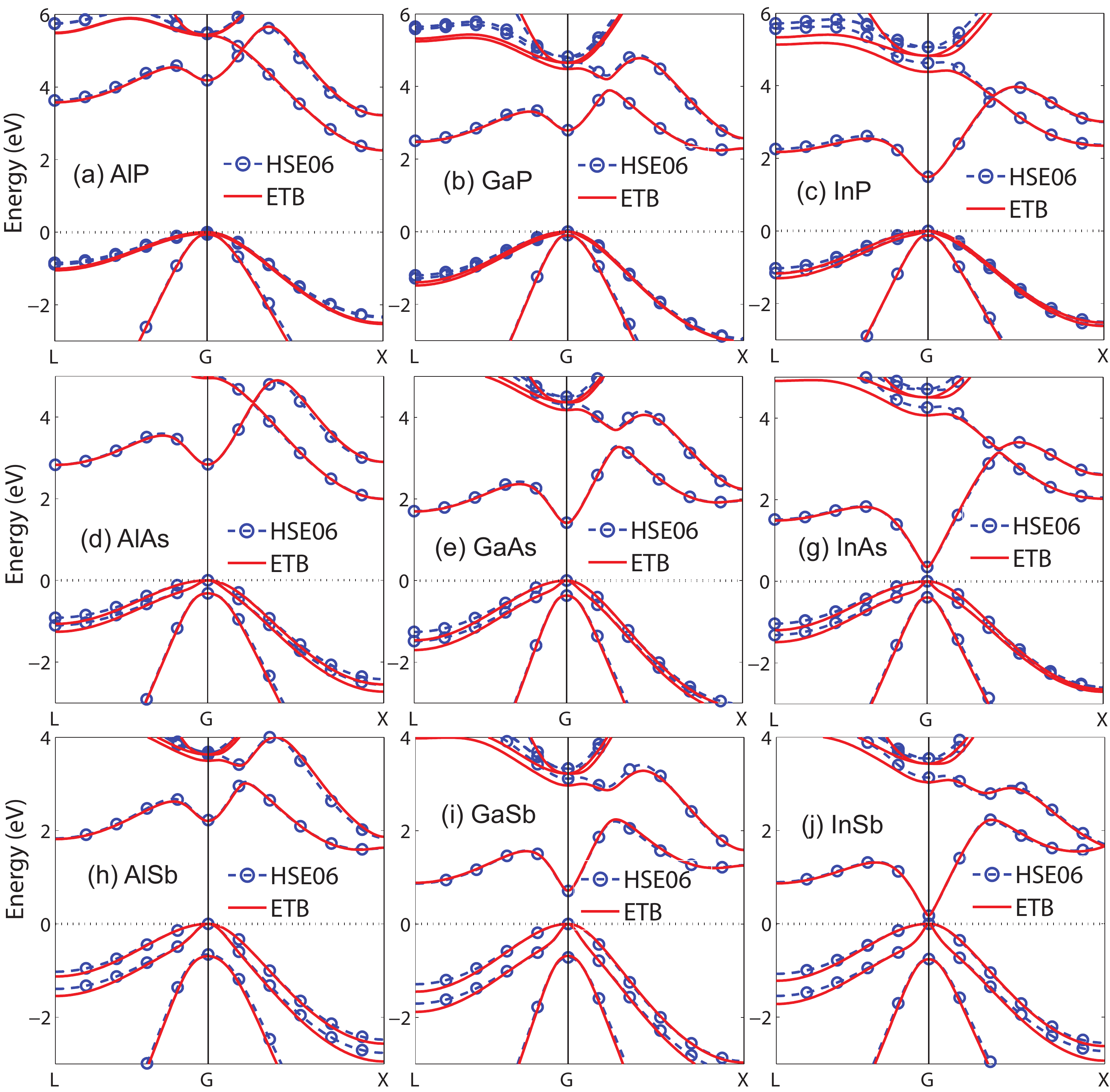}
\caption{Band structure of III-Vs materials with ETB and HSE06
calculation. Presented band structures of III-V materials include
(a) AlP , (b) GaP , (c) InP ,(d) AlAs , (e) GaAs,(f) InAs ,(g) AlSb
,(h) GaSb ,(i) InSb. ETB band structures are in good agreement with
HSE06 results. }\label{fig_IIIVs}
\end{figure*}

\begin{table*}
    \small
    \setlength{\tabcolsep}{1pt}
\begin{tabular}
[c]{c|c|c|c}\hline\hline
& AlP & GaP & InP \\\hline%
\begin{tabular}
[c]{c}%
targets\\\hline
$E_{g}(\Gamma)$\\
$E_{g}(X)$\\
$E_{g}(L)$\\
$\Delta_{SO}$\\
\\
$m_{hh100}$\\
$m_{hh110}$\\
$m_{hh111}$\\
$m_{lh100}$\\
$m_{lh110}$\\
$m_{lh111}$\\
$m_{so}$\\
\\
$m_{c\Gamma}$\\
$m_{cXl}$\\
$m_{cXt}$\\
$m_{cLl}$\\
$m_{cLt}$\\
\end{tabular}
&
\begin{tabular}
[c]{cc}%
\begin{tabular}
[c]{ccc}%
HSE06 & ETB & error \\\hline
		  $4.305$ &           $4.303$ &                $0.0\%$ \\
          $2.391$ &           $2.327$ &                $2.0\%$ \\
          $3.751$ &           $3.715$ &                $1.0\%$ \\
          $0.064$ &           $0.064$ &                $0.0\%$ \\
  &  & \\
             $0.508$ &              $0.505$ &              $0.7\%$ \\ 
             $0.998$ &              $0.981$ &              $1.7\%$ \\ 
             $1.273$ &              $1.270$ &              $0.2\%$ \\ 
             $0.250$ &              $0.237$ &              $5.0\%$ \\ 
             $0.201$ &              $0.193$ &              $3.9\%$ \\ 
             $0.193$ &              $0.185$ &              $4.0\%$ \\ 
             $0.343$ &              $0.328$ &              $4.3\%$ \\ 
  &  & \\\ 
             $0.189$ &              $0.185$ &              $2.4\%$ \\ 
             $0.781$ &              $0.789$ &              $1.0\%$ \\ 
             $0.242$ &              $0.231$ &              $4.8\%$ \\ 
             $1.610$ &              $1.674$ &              $3.9\%$ \\ 
             $0.177$ &              $0.192$ &              $8.6\%$ \\
\end{tabular}
&
\begin{tabular}
[c]{c}%
Ref \\
\hline
                        $3.55$ \\ 
                        $2.49$ \\ 
                        $3.54$ \\ 
                        $0.07$ \\ 
 \\ 
                        $0.52$ \\ 
                        $0.87$ \\ 
                        $1.12$ \\ 
                        $0.21$ \\ 
                        $0.18$ \\ 
                        $0.17$ \\ 
                        $0.30$ \\
                        \\ 
                        $0.22$ \\ 
                        $2.68$ \\ 
                        $0.16$ \\ 
                       $-$ \\ 
                       $-$ \\ 
\end{tabular}
\\
\end{tabular}
&
\begin{tabular}
[c]{cc}%
\begin{tabular}
[c]{ccc}%
HSE06 & ETB & error\\\hline
          $2.797$ &           $2.793$ &                $0.1\%$ \\
          $2.256$ &           $2.250$ &                $0.3\%$ \\
          $2.504$ &           $2.492$ &                $0.5\%$ \\
          $0.098$ &           $0.098$ &                $0.0\%$ \\
  &  & \\             
             $0.355$ &              $0.351$ &              $1.4\%$ \\ 
             $0.667$ &              $0.655$ &              $1.9\%$ \\ 
             $0.843$ &              $0.836$ &              $0.9\%$ \\ 
             $0.160$ &              $0.153$ &              $4.0\%$ \\ 
             $0.132$ &              $0.127$ &              $3.4\%$ \\ 
             $0.127$ &              $0.122$ &              $3.5\%$ \\ 
             $0.229$ &              $0.222$ &              $3.3\%$ \\ 
   &  & \\  
             $0.131$ &              $0.132$ &              $0.5\%$ \\ 
             $1.532$ &              $1.305$ &             $14.8\%$ \\ 
             $0.224$ &              $0.231$ &              $3.0\%$ \\ 
             $1.581$ &              $1.722$ &              $8.9\%$ \\ 
             $0.138$ &              $0.163$ &             $18.2\%$ \\ 
\end{tabular}
&
\begin{tabular}
[c]{c}%
Ref \\
\hline
                        $2.89$ \\ 
                        $2.28$ \\ 
                        $2.64$ \\ 
                        $0.08$ \\ 
 \\ 
                        $0.33$ \\ 
                        $0.52$ \\ 
                        $0.65$ \\ 
                        $0.20$ \\ 
                        $0.16$ \\ 
                        $0.15$ \\ 
                        $0.25$ \\
                        \\ 
                        $0.13$ \\ 
                        $2.00$ \\ 
                        $0.25$ \\ 
                        $1.20$ \\ 
                        $0.15$ \\
\end{tabular}
\\
\end{tabular}
&
\begin{tabular}
[c]{cc}%
\begin{tabular}
[c]{ccc}%
HSE06 & ETB & error\\\hline
          $1.397$ &           $1.391$ &                $0.4\%$ \\
          $2.283$ &           $2.272$ &                $0.4\%$ \\
          $2.162$ &           $2.143$ &                $0.9\%$ \\
          $0.124$ &           $0.124$ &                $0.0\%$ \\
  &  & \\             
             $0.405$ &              $0.403$ &              $0.4\%$ \\ 
             $0.726$ &              $0.728$ &              $0.2\%$ \\ 
             $0.918$ &              $0.942$ &              $2.6\%$ \\ 
             $0.114$ &              $0.110$ &              $3.2\%$ \\ 
             $0.102$ &              $0.098$ &              $3.1\%$ \\ 
             $0.099$ &              $0.095$ &              $3.1\%$ \\ 
             $0.190$ &              $0.186$ &              $1.8\%$ \\
  &  & \\               
             $0.087$ &              $0.084$ &              $3.4\%$ \\ 
             $1.476$ &              $1.348$ &              $8.6\%$ \\ 
             $0.244$ &              $0.251$ &              $2.6\%$ \\ 
             $1.984$ &              $1.941$ &              $2.2\%$ \\ 
             $0.144$ &              $0.166$ &             $15.5\%$ \\
\end{tabular}
&
\begin{tabular}
[c]{c}%
Ref \\
\hline
                        $3.00$ \\ 
                        $2.38$ \\ 
                        $1.94$ \\ 
                        $0.11$ \\ 
 \\ 
                        $0.53$ \\ 
                        $0.88$ \\ 
                        $1.14$ \\ 
                        $0.12$ \\ 
                        $0.11$ \\ 
                        $0.11$ \\ 
                        $0.21$ \\
                        \\ 
                        $0.08$ \\ 
                       $-$ \\ 
                       $-$ \\ 
                       $-$ \\ 
                       $-$ \\ 
\end{tabular}
\\
\end{tabular}
\\\hline
\end{tabular}
\centering \caption{Targets comparison of bulk XP. Critical band
edges and effective masses at $\Gamma$, $X$ and $L$ from TB and
HSE06 calculations are compared. The $E_{g}$ and $\Delta_{SO}$ are
in the unit of eV; effective masses are scaled by free electron mass
$m_0$. The error column summarizes the relative discrepancies between HSE06
and ETB results. The Reference bandedge and effective masses are from Ref \onlinecite{IIIV_para_Vurgulfman}.
}%
\label{tab:targets_comparison_XP} %
\end{table*}

\begin{table*}
    \small
    \setlength{\tabcolsep}{1pt}
\begin{tabular}
[c]{c|c|c|c}\hline\hline
& AlAs & GaAs & InAs \\\hline%
\begin{tabular}
[c]{c}%
targets\\\hline
$E_{g}(\Gamma)$\\
$E_{g}(X)$\\
$E_{g}(L)$\\
$\Delta_{SO}$\\
\\
$m_{hh100}$\\
$m_{hh110}$\\
$m_{hh111}$\\
$m_{lh100}$\\
$m_{lh110}$\\
$m_{lh111}$\\
$m_{so}$ \\
\\
$m_{c\Gamma}$\\
$m_{cXl}$\\
$m_{cXt}$\\
$m_{cLl}$\\
$m_{cLt}$\\
\end{tabular}
&
\begin{tabular}
[c]{cc}%
\begin{tabular}
[c]{ccc}%
HSE06 & ETB & error \\\hline
          $2.891$ &           $2.887$ &                $0.2\%$ \\
          $2.050$ &           $2.054$ &                $0.2\%$ \\
          $2.880$ &           $2.872$ &                $0.3\%$ \\
          $0.317$ &           $0.317$ &                $0.0\%$ \\
  &  & \\             
             $0.437$ &              $0.441$ &              $1.0\%$ \\ 
             $0.838$ &              $0.841$ &              $0.4\%$ \\ 
             $1.082$ &              $1.104$ &              $2.0\%$ \\ 
             $0.166$ &              $0.161$ &              $2.9\%$ \\ 
             $0.141$ &              $0.137$ &              $2.3\%$ \\ 
             $0.135$ &              $0.132$ &              $2.4\%$ \\ 
             $0.272$ &              $0.257$ &              $5.6\%$ \\ 
   &  & \\
             $0.126$ &              $0.123$ &              $2.2\%$ \\ 
             $0.850$ &              $0.864$ &              $1.6\%$ \\ 
             $0.231$ &              $0.223$ &              $3.5\%$ \\ 
             $1.557$ &              $1.627$ &              $4.5\%$ \\ 
             $0.144$ &              $0.160$ &             $10.6\%$ \\ 
\end{tabular}
&
\begin{tabular}
[c]{c}%
Ref \\
\hline
                        $3.00$ \\ 
                        $2.16$ \\ 
                        $2.35$ \\ 
                        $0.34$ \\ 
 \\ 
                        $0.47$ \\ 
                        $0.82$ \\ 
                        $1.09$ \\ 
                        $0.19$ \\ 
                        $0.16$ \\ 
                        $0.15$ \\ 
                        $0.28$ \\
                        \\ 
                        $0.15$ \\ 
                        $0.97$ \\ 
                        $0.22$ \\ 
                        $1.32$ \\ 
                        $0.15$ \\ 
\end{tabular}
\\
\end{tabular}
&
\begin{tabular}
[c]{cc}%
\begin{tabular}
[c]{ccc}%
HSE06 & ETB & error \\\hline
          $1.418$ &           $1.416$ &                $0.2\%$ \\
          $1.919$ &           $1.912$ &                $0.4\%$ \\
          $1.701$ &           $1.692$ &                $0.6\%$ \\
          $0.367$ &           $0.367$ &                $0.0\%$ \\
  &  & \\
             $0.308$ &              $0.317$ &              $3.0\%$ \\ 
             $0.569$ &              $0.581$ &              $2.2\%$ \\ 
             $0.744$ &              $0.762$ &              $2.4\%$ \\ 
             $0.081$ &              $0.081$ &              $0.8\%$ \\ 
             $0.073$ &              $0.072$ &              $0.3\%$ \\ 
             $0.070$ &              $0.070$ &              $0.2\%$ \\ 
             $0.162$ &              $0.156$ &              $3.8\%$ \\ 
   &  & \\
             $0.065$ &              $0.066$ &              $1.3\%$ \\ 
             $1.564$ &              $1.331$ &             $14.9\%$ \\ 
             $0.213$ &              $0.216$ &              $1.4\%$ \\ 
             $1.613$ &              $1.669$ &              $3.5\%$ \\ 
             $0.110$ &              $0.129$ &             $17.9\%$ \\ 
\end{tabular}
&
\begin{tabular}
[c]{c}%
Ref \\
\hline
                        $1.42$ \\ 
                        $1.90$ \\ 
                        $1.70$ \\ 
                        $0.28$ \\ 
 \\ 
                        $0.35$ \\ 
                        $0.64$ \\ 
                        $0.89$ \\ 
                        $0.09$ \\ 
                        $0.08$ \\ 
                        $0.08$ \\ 
                        $0.17$ \\ 
                        \\
                        $0.07$ \\ 
                        $1.30$ \\ 
                        $0.23$ \\ 
                        $1.90$ \\ 
                        $0.08$ \\
\end{tabular}
\\
\end{tabular}
&
\begin{tabular}
[c]{cc}%
\begin{tabular}
[c]{ccc}%
HSE06 & ETB & error \\\hline
          $0.350$ &           $0.348$ &                $0.7\%$ \\
          $2.052$ &           $2.021$ &                $1.5\%$ \\
          $1.514$ &           $1.502$ &                $0.8\%$ \\
          $0.391$ &           $0.391$ &                $0.0\%$ \\
  &  & \\            
             $0.344$ &              $0.352$ &              $2.2\%$ \\ 
             $0.625$ &              $0.639$ &              $2.3\%$ \\ 
             $0.835$ &              $0.865$ &              $3.6\%$ \\ 
             $0.026$ &              $0.026$ &              $1.0\%$ \\ 
             $0.026$ &              $0.026$ &              $1.0\%$ \\ 
             $0.025$ &              $0.025$ &              $0.9\%$ \\ 
             $0.102$ &              $0.095$ &              $6.7\%$ \\ 
   &  & \\    
             $0.022$ &              $0.021$ &              $1.6\%$ \\ 
             $1.458$ &              $1.275$ &             $12.5\%$ \\ 
             $0.232$ &              $0.238$ &              $2.4\%$ \\ 
             $1.904$ &              $1.820$ &              $4.4\%$ \\ 
             $0.114$ &              $0.131$ &             $15.0\%$ \\ 
\end{tabular}
&
\begin{tabular}
[c]{c}%
Ref \\
\hline
                        $0.35$ \\ 
                        $1.37$ \\ 
                        $1.07$ \\ 
                        $0.39$ \\ 
 \\ 
                        $0.33$ \\ 
                        $0.51$ \\ 
                        $0.62$ \\ 
                        $0.03$ \\ 
                        $0.03$ \\ 
                        $0.03$ \\ 
                        $0.14$ \\
                        \\ 
                        $0.03$ \\ 
                        $1.13$ \\ 
                        $0.16$ \\ 
                        $0.64$ \\ 
                        $0.05$ \\ 
\end{tabular}
\\
\end{tabular}
\\\hline
\end{tabular}
\centering \caption{Targets comparison of bulk XAs. Critical band
edges and effective masses at $\Gamma$, $X$ and $L$ from TB and
HSE06 calculations are compared. The $E_{g}$ and $\Delta_{SO}$ are
in the unit of eV; effective masses are scaled by free electron mass
$m_0$. The error column summarizes the relative discrepancies between HSE06
and ETB results. The Reference bandedge and effective masses are from Ref \onlinecite{IIIV_para_Vurgulfman}.
}%
\label{tab:targets_comparison_XAs} %
\end{table*}

\begin{table*}
    \small
    \setlength{\tabcolsep}{1pt}
\begin{tabular}
[c]{c|c|c|c}\hline\hline
& AlSb & GaSb & InSb \\\hline%
\begin{tabular}
[c]{c}%
targets\\\hline
$E_{g}(\Gamma)$\\
$E_{g}(X)$\\
$E_{g}(L)$\\
$\Delta_{SO}$\\
\\
$m_{hh100}$\\
$m_{hh110}$\\
$m_{hh111}$\\
$m_{lh100}$\\
$m_{lh110}$\\
$m_{lh111}$\\
$m_{so}$\\
\\
$m_{c\Gamma}$\\
$m_{cXl}$\\
$m_{cXt}$\\
$m_{cLl}$\\
$m_{cLt}$\\
\end{tabular}
&
\begin{tabular}
[c]{cc}%
\begin{tabular}
[c]{ccc}%
HSE06 & ETB & error \\\hline
          $2.223$ &           $2.225$ &                $0.1\%$ \\
          $1.597$ &           $1.601$ &                $0.2\%$ \\
          $1.831$ &           $1.835$ &                $0.2\%$ \\
          $0.655$ &           $0.642$ &                $1.9\%$ \\
  &  & \\
             $0.315$ &              $0.322$ &              $2.4\%$ \\ 
             $0.593$ &              $0.615$ &              $3.6\%$ \\ 
             $0.761$ &              $0.805$ &              $5.8\%$ \\ 
             $0.125$ &              $0.121$ &              $3.4\%$ \\ 
             $0.106$ &              $0.103$ &              $2.8\%$ \\ 
             $0.102$ &              $0.099$ &              $2.8\%$ \\ 
             $0.238$ &              $0.220$ &              $7.7\%$ \\ 
  &  & \\ 
             $0.108$ &              $0.109$ &              $1.0\%$ \\ 
             $1.458$ &              $1.216$ &             $16.6\%$ \\ 
             $0.219$ &              $0.209$ &              $4.7\%$ \\ 
             $1.520$ &              $1.543$ &              $1.5\%$ \\ 
             $0.121$ &              $0.132$ &              $8.9\%$ \\  
\end{tabular}
&
\begin{tabular}
[c]{c}%
Ref \\
\hline
                        $2.30$ \\ 
                        $1.62$ \\ 
                        $2.21$ \\ 
                        $0.68$ \\ 
 \\ 
                        $0.36$ \\ 
                        $0.61$ \\ 
                        $0.81$ \\ 
                        $0.13$ \\ 
                        $0.11$ \\ 
                        $0.11$ \\ 
                        $0.22$ \\ 
                        \\
                        $0.14$ \\ 
                        $1.36$ \\ 
                        $0.12$ \\ 
                        $1.64$ \\ 
                        $0.23$ \\ 
\end{tabular}
\\
\end{tabular}
&
\begin{tabular}
[c]{cc}%
\begin{tabular}
[c]{ccc}%
HSE06 & ETB & error \\\hline
          $0.707$ &           $0.703$ &                $0.5\%$ \\
          $1.205$ &           $1.202$ &                $0.2\%$ \\
          $0.865$ &           $0.870$ &                $0.6\%$ \\
          $0.714$ &           $0.714$ &                $0.0\%$ \\
  &  & \\             
             $0.232$ &              $0.251$ &              $8.3\%$ \\ 
             $0.426$ &              $0.456$ &              $7.0\%$ \\ 
             $0.566$ &              $0.606$ &              $7.0\%$ \\ 
             $0.041$ &              $0.041$ &              $0.8\%$ \\ 
             $0.038$ &              $0.038$ &              $0.1\%$ \\ 
             $0.037$ &              $0.037$ &              $0.1\%$ \\ 
             $0.137$ &              $0.124$ &              $9.5\%$ \\ 
  &  & \\   
             $0.037$ &              $0.037$ &              $0.3\%$ \\ 
             $2.362$ &              $1.826$ &             $22.7\%$ \\ 
             $0.194$ &              $0.219$ &             $12.5\%$ \\ 
             $1.587$ &              $1.568$ &              $1.2\%$ \\ 
             $0.090$ &              $0.108$ &             $19.2\%$ \\ 
\end{tabular}
&
\begin{tabular}
[c]{c}%
Ref \\
\hline
                        $0.73$ \\ 
                        $1.03$ \\ 
                        $0.75$ \\ 
                        $0.76$ \\ 
 \\ 
                        $0.25$ \\ 
                        $0.49$ \\ 
                        $0.71$ \\ 
                        $0.04$ \\ 
                        $0.04$ \\ 
                        $0.04$ \\ 
                        $0.12$ \\ 
                        \\
                        $0.04$ \\ 
                        $1.51$ \\ 
                        $0.22$ \\ 
                        $1.30$ \\ 
                        $0.10$ \\
\end{tabular}
\\
\end{tabular}
&
\begin{tabular}
[c]{cc}%
\begin{tabular}
[c]{ccc}%
HSE06 & ETB & error \\\hline
          $0.172$ &           $0.170$ &                $1.6\%$ \\ 
          $1.566$ &           $1.549$ &                $1.1\%$ \\ 
          $0.891$ &           $0.867$ &                $2.8\%$ \\ 
          $0.754$ &           $0.770$ &                $0.7\%$ \\ 
  &  & \\             
             $0.245$ &              $0.277$ &             $12.9\%$ \\ 
             $0.452$ &              $0.507$ &             $12.2\%$ \\ 
             $0.609$ &              $0.694$ &             $13.9\%$ \\ 
             $0.012$ &              $0.013$ &              $6.1\%$ \\ 
             $0.013$ &              $0.014$ &              $4.7\%$ \\ 
             $0.012$ &              $0.012$ &              $6.6\%$ \\ 
             $0.117$ &              $0.108$ &              $7.5\%$ \\ 
  &  & \\   
             $0.011$ &              $0.012$ &              $8.7\%$ \\ 
             $0.877$ &              $0.790$ &             $10.0\%$ \\ 
             $0.219$ &              $0.230$ &              $5.0\%$ \\ 
             $1.685$ &              $1.575$ &              $6.5\%$ \\ 
             $0.096$ &              $0.111$ &             $15.7\%$ \\ 
\end{tabular}
&
\begin{tabular}
[c]{c}%
Ref \\
\hline
                         $0.17$ \\ 
                       $-$ \\ 
                        $0.93$ \\ 
                        $0.81$ \\ 
 \\ 
                        $0.26$ \\ 
                        $0.43$ \\ 
                        $0.56$ \\ 
                        $0.02$ \\ 
                        $0.01$ \\ 
                        $0.01$ \\ 
                        $0.11$ \\ 
                        \\
                        $0.01$ \\ 
                       $-$ \\ 
                       $-$ \\ 
                       $-$ \\ 
                       $-$ \\
\end{tabular}
\\
\end{tabular}
\\\hline
\end{tabular}
\centering \caption{Targets comparison of bulk XSb. Critical band
edges and effective masses at $\Gamma$, $X$ and $L$ from TB and
HSE06 calculations are compared. The $E_{g}$ and $\Delta_{SO}$ are
in the unit of eV; effective masses are scaled by free electron mass
$m_0$. The error column summarizes the relative discrepancies between HSE06
and ETB results.  The Reference bandedge and effective masses are from Ref \onlinecite{IIIV_para_Vurgulfman}.
}%
\label{tab:targets_comparison_XSb} %
\end{table*}

\begin{figure*}
\centering
\includegraphics[width=\textwidth]{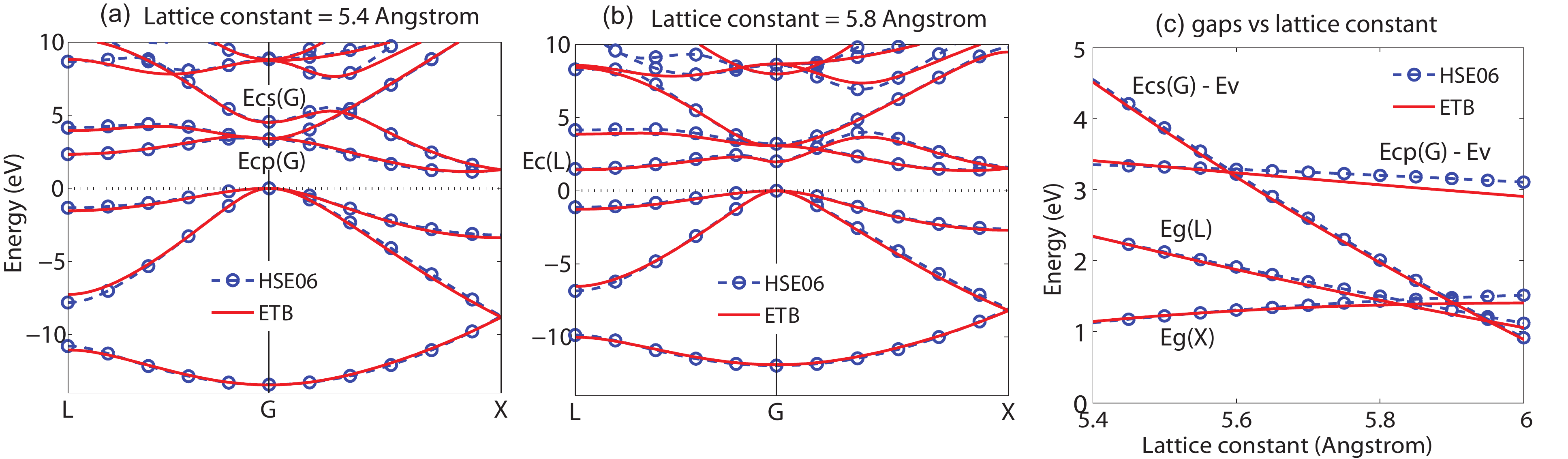}
\caption{Band structure of Si with different lattice constants. (a)
Si with a lattice constant of 5.4 ${\AA}$, (b)Si with a lattice
constant of 5.8 ${\AA}$, (c) direct and indirect band gaps of Si
with different lattice constants. The lowest conduction band at
$\Gamma$ point transit from p-bands to s-bands at about 5.8 ${\AA}$.
When lattice constant is 5.4 ${\AA}$, Si is a indirect gap
semiconductor, the X conduction valley is the lowest conduction
valley. As lattice constant increases, the band gap at of X valley
(Eg(X)) increases slightly, while the bandgap of L valleys (Eg(L))
and direct band gap ($E_{cs}(G)-E_v$) decrease significantly.
}\label{fig_Si_hydro_static}
\end{figure*}
\section{results}\label{sec:results}
In this work, \textit{ab-initio} level calculations of group IV and
III-V systems are performed with VASP \cite{VASP_Kresse}. The
screened hybrid functional of Heyd, Scuseria, and Ernzerhof
(HSE06)\cite{HSE06_original} is used to produce the bulk and the
superlattices band structures with band gaps
 comparable with experiments\cite{Vasp_HSE}. In the HSE06 hybrid functional method scheme,
the total exchange energy incorporates 25\% short-range Hartree-Fock
(HF) exchange and 75\% Perdew-Burke-Ernzerhof(PBE)
exchange\cite{PBE}. The screening parameter $\mu$ which defines the
range separation is empirically set to 0.2 $\AA$ for both the HF and
PBE parts. The correlation energy is described by the PBE
functional. In all presented HSE06 calculations, a cutoff energy of
350eV is used. $\Gamma$-point centered Monkhorst Pack kspace grids
are used for both bulk and superlattice systems. The size of the
kspace grid for strained bulk calculations is $6\times6\times6$,
while one for 001 superlattices is $6\times6\times3$. k-points with
integration weights equal to zero are added to the original uniform
grids in order to generate energy bands with higher k-space
resolution. PAW\cite{VASP_PP} pseudopotentials are used in all HSE06
calculations. The pseudopotentials for all atoms include the
outermost occupied s and p atomic states as valence states.
\textit{Ab-initio} band structures of strained and unstrained bulk
materials are aligned based on model solid
theory\cite{VandeWalle_modelsolid_IIIV,VandeWalle_modelsolid_SiGe}.
With the model solid theory, relative band offsets are determined by
using different superlattices. 

\begin{table*}
    \small
\setlength{\tabcolsep}{1pt}
\begin{tabular}
[c]{c|c|c}\hline\hline
& Si & Ge \\\hline%
\begin{tabular}
[c]{c}%
targets \\
$b_v$\\
$d_v$\\
$\Xi_{001}$\\
$\Xi_{110}$\\
\end{tabular}
&
\begin{tabular}
[c]{cc}%
\begin{tabular}
[c]{ccc}%
HSE06 & ETB & error \\              
              $2.58$ &               $2.60$ &              $0.8\%$ \\ 
              $6.01$ &               $5.78$ &              $3.8\%$ \\ 
              $8.31$ &               $8.23$ &              $1.0\%$ \\ 
             $15.59$ &              $15.22$ &              $2.4\%$ \\ 
\end{tabular}
& 
\begin{tabular}
[c]{c}%
Ref \\              
              $2.10$ \\ 
              $4.85$ \\ 
              $8.60$ \\ 
              $-$  \\ 
\end{tabular}
 \\
\end{tabular}
&
\begin{tabular}
[c]{cc}
\begin{tabular}
[c]{ccc}
HSE06 & ETB & error \\
              $2.81$ &               $2.80$ &              $0.1\%$ \\ 
              $5.88$ &               $5.89$ &              $0.0\%$ \\ 
              $8.35$ &               $8.35$ &              $0.0\%$ \\ 
             $17.21$ &              $17.10$ &              $0.6\%$ \\ 
\end{tabular}
& 
\begin{tabular}
[c]{c}%
Ref \\              
              $2.86$ \\ 
              $5.28$ \\ 
              $-$ \\ 
              $-$  \\ 
\end{tabular}
\\
\end{tabular}
\\\hline
\end{tabular}
\centering \caption{Targets comparison of deformation potentials of
group IV materials. Reference experimental values are from Ref. \onlinecite{VandeWalle_modelsolid_SiGe}.
}%
\label{tab:strained_targets_comparison_Ge_Si} %
\end{table*}

\begin{figure}[h]\centering
\includegraphics[width=0.5\textwidth]{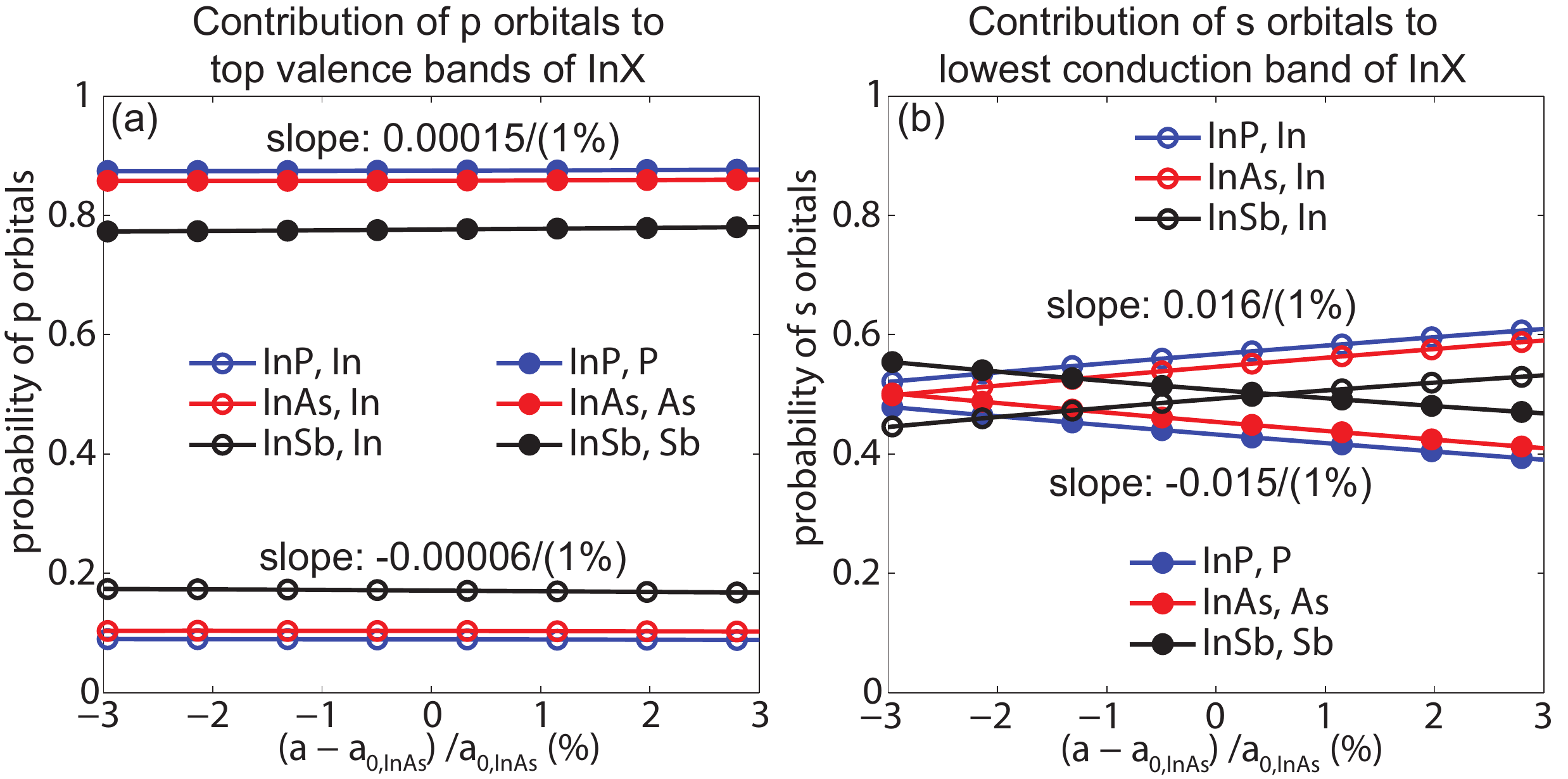}
\caption{ Contribution of p orbitals to the top valence bands(a) and
contribution of s orbitals to the lowest conduction bands of InX (X=P,As,Sb).
The p orbitals of In and cation atoms contribute to the top valence
bands. When lattice constant change one percent, p orbitals
contribution are changed by less than 0.0002. The s orbitals of In
and anion atoms contribute to the lowest conduction bands. When lattice
constant change one percent, s orbitals contribution are changed by
less than 0.02. }\label{fig_strain_wf}
\end{figure}
\subsection{Room temperature targets}
\textit{Ab-initio} calculations usually assume zero temperature, while ETB models matching room temperature experiments are required for realistic device modeling. In this work, in order to get \textit{ab-initio} band structures
matching experiments under room temperature, artificial hydrostatic
strain is applied to individual material to mimic the effect of room
temperature and to compensate the error of \textit{ab-initio}
calculations. With hydrostatic strain, lattice constants change from
$a_0$ to $a_0+\delta a$. This artificial lattice constant change can
be used to adjust the \textit{ab-initio} band gap of semiconductors
to match finite temperature experimental band gap. Table
\ref{tab:change of lattice const} shows the required $\delta a$ in
order to match HSE06 band gaps with room temperature experimental
data. It can be seen that the most of the required $\delta a$ are in
general less than $1\%$ hydrostatic strain. The AlP requires $\delta
a$ up to $2\% a_0$. By this adjustment, band gaps of most of the
presented semiconductors reach less than 0.05eV mismatch compared
with experimental results. The largest mismatch appears in AlAs
which has the mismatch of about 0.1eV.
\begin{table}[h]
\centering \small \setlength{\tabcolsep}{3pt}
\begin{tabular}{c|c|c|c|c|c}
\hline
material & $a_0$ ($\AA$)  & gap (eV)  & $\delta a$ ($\AA$) & $\delta a/a_0$($\%$) & gap (eV) \\
& exp,300K & exp,300K & HSE06 & HSE06 & HSE06\\
\hline
Si & 5.43 & 1.12 & -0.0273 & 0.5 & 1.141 \\
Ge & 5.658 & 0.66 & -0.010 & -0.2 &0.755 \\
AlP & 5.4672 & 2.488 & 0.124 & 2.3 & 2.391 \\ 
GaP & 5.4505 & 2.273 & 0.01 & 0.2 & 2.256 \\
InP & 5.8697 & 1.353 & 0.042 & 0.7 & 1.397 \\ 
AlAs & 5.6611 & 2.164 & 0.05 & 0.9 & 2.05 \\
GaAs & 5.6533 & 1.422 & -0.0226 & -0.4 & 1.418 \\
InAs & 6.0583 & 0.354 & 0.0221  & 0.4 & 0.350 \\
AlSb & 6.1355 & 1.616 & -0.0186 & 0.3 & 1.597 \\
GaSb & 6.0959 & 0.727 & -0.0045 & -0.1 & 0.707 \\
InSb & 6.4794 & 0.174 & 0.0406  & 0.6 & 0.172 \\
\hline
\end{tabular}
\caption{Experimental lattice constants and band gaps of group IV
and III-V materials under room temperature; required changes of
lattice constants $\delta a$ in order to match HSE06 band gap with
experiments.}\label{tab:change of lattice const}
\end{table}
Since the parameterization algorithm used in this work relies on the
\textit{ab-initio} wave functions, the concern of this artificial
adjustment is that whether it will change \textit{ab-initio} wave
functions significantly. Fig. \ref{fig_strain_wf} shows the
contribution of different orbitals in \textit{ab-initio} wave
functions as a function of lattice constant. Here the
\textit{ab-initio} wave functions of InX with different lattice
constants are represented by the same basis functions. It can be
seen that the every percent of hydrostatic strain introduced changes
the contribution of orbitals up to 0.02. Thus the artificial
adjustment introduces negligible changes to wave functions.
Similar trend can be observed in other
group III-V and IV materials.
In this work, the ETB parameters are all fitted with respect to
\textit{ab-initio} results that are adjusted with respect to room
temperature experiments.

\subsection{ETB parameters for strained materials}\label{sec:parameters}
The ETB model in this work makes use of sp3d5s* basis functions. The
sp3d5s* empirical ETB model with nearest neighbor interactions has
been proved to be a sufficient model for bulk zincblende and diamond
structures\cite{Boykin_SiGe, Boykin_TB_strain,
YaohuaTan_abinitioMapping}. To parameterize the ETB model from
\textit{ab-initio} results, both \textit{ab-initio} band structure
and wave functions are considered as fitting targets. The process of
parameterization from \textit{ab-initio} results was described by
Ref. \onlinecite{YaohuaTan_abinitioMapping}.
This method is applicable to any model that is able to deliver explicit wave 
functions, and is not restricted to the HSE06 calculations. E.g. empirical pseudopotential calculations or more expensive but accurate GW calculations
can be used.

To obtain ETB parameters for strained materials, the process of
parameterization from \textit{ab-initio} results by Ref.
\onlinecite{YaohuaTan_abinitioMapping} is applied to multiple
strained systems. To consider multiple systems in the fitting
process, a total fitness to be minimized is defined as a summation
of fitness of all systems considered (labeled by index $s$)
$F_{total} = \sum_{s} F_{s} $. The fitness $F_{s}$ is defined to
capture important targets of each stained system considered in the
fitting process. The strained systems considered in this work are
shown by Fig. \ref{fig_strain_displacement}, including zincblende or
diamond structures with a) hydro static strain, b) pure bond length
changes, c) diagonal strains and d) off-diagonal strain. For
Hydrostatic strain cases, materials with different lattice constant
ranging from 5.2 to 6.6 $\AA$ are considered. While for other kind
of strains, strains with amplitudes from $-4 \%$ to $4 \%$ are
considered.

For hydrostatically strained materials, fitting targets includes
band structures, important band edges, effective masses and wave
functions at high symmetry points. Those targets were considered in
previous work (ref.\onlinecite{YaohuaTan_abinitioMapping}) in order
to get ETB parameters for unstrained bulk materials. To extract ETB
parameters for arbitrarily strained materials, wave functions and
energies at high symmetry points are also considered as fitting
targets. For strained systems, it is sufficient to use the strain
induced band edge splitting at high symmetry points as targets.
Effective masses at those points are not considered as fitting
targets. Effective masses in strained materials are related to the
splitting of band edges and effective masses of unstrained systems.
For example, the effective masses of valence bands in a strained
group III-V or IV material can be well described by a Luttinger
model \cite{8bandkdotp_Bahder}. The well known conduction band
effective mass change under shear strain( with strain component
$\epsilon_{xy}$ ) can also described by camel back
model\cite{YaohuaTan_kp_strain}. Those models include the strain
effect as k-independent perturbation terms. The strain induced terms
correspond to the band edge splitting at high symmetry points.

It should be noted that the usage of wave function data eliminates the 
arbitrariness of parameters among materials. It can be seen from tables
\ref{tab:onsite_soc},\ref{tab:interatomic_parameters_bond_length_SiGe IIIVs}, \ref{tab:off_diagonal_onsite_parameters_multipole},\ref{tab:interatomic_coupling_parameters_multipole}
that the parameters of different materials have small relative variations. 
Many of the tight binding parameters show a clear monotonic dependence of the
principle quantum number of atoms. For instance, the $V_{p_cp_a\sigma}$'s have a 
trend $|V_{p_cp_P\sigma}|<|V_{p_cp_{As}\sigma}|<|V_{p_cp_{Sb}\sigma}|$ as it is shown 
in table \ref{tab:interatomic_parameters_bond_length_SiGe IIIVs}. 
This trend of parameters is related to
the wave functions of top valence bands at $\Gamma$ point.
Similar to the trend of $V_{p_cp_a\sigma}$, 
the contribution of $p$ orbitals of cations $w_{p_c}$ also shows a monotonic trend 
of $w_{p_c}(InP)<w_{p_c}(InAs)<w_{p_c}(InSb)$, while the $p$ orbitals of anions $w_{p_a}$ 
show an opposite trend $w_{p_a}(InP)>w_{p_a}(InAs)>w_{p_a}(InSb)$ as it is shown 
in Fig. \ref{fig_strain_wf} (a).
Furthermore, the Fig. \ref{fig_strain_wf} also shows  that the InX orbitals have a 
similar rate of variation under hydrostatic strain; consequently,  
the scaling factor $\eta_{pp\sigma}$'s for all materials has the value from $0.94$ to $1.05$.

\begin{figure}
\includegraphics[width=0.5\textwidth]{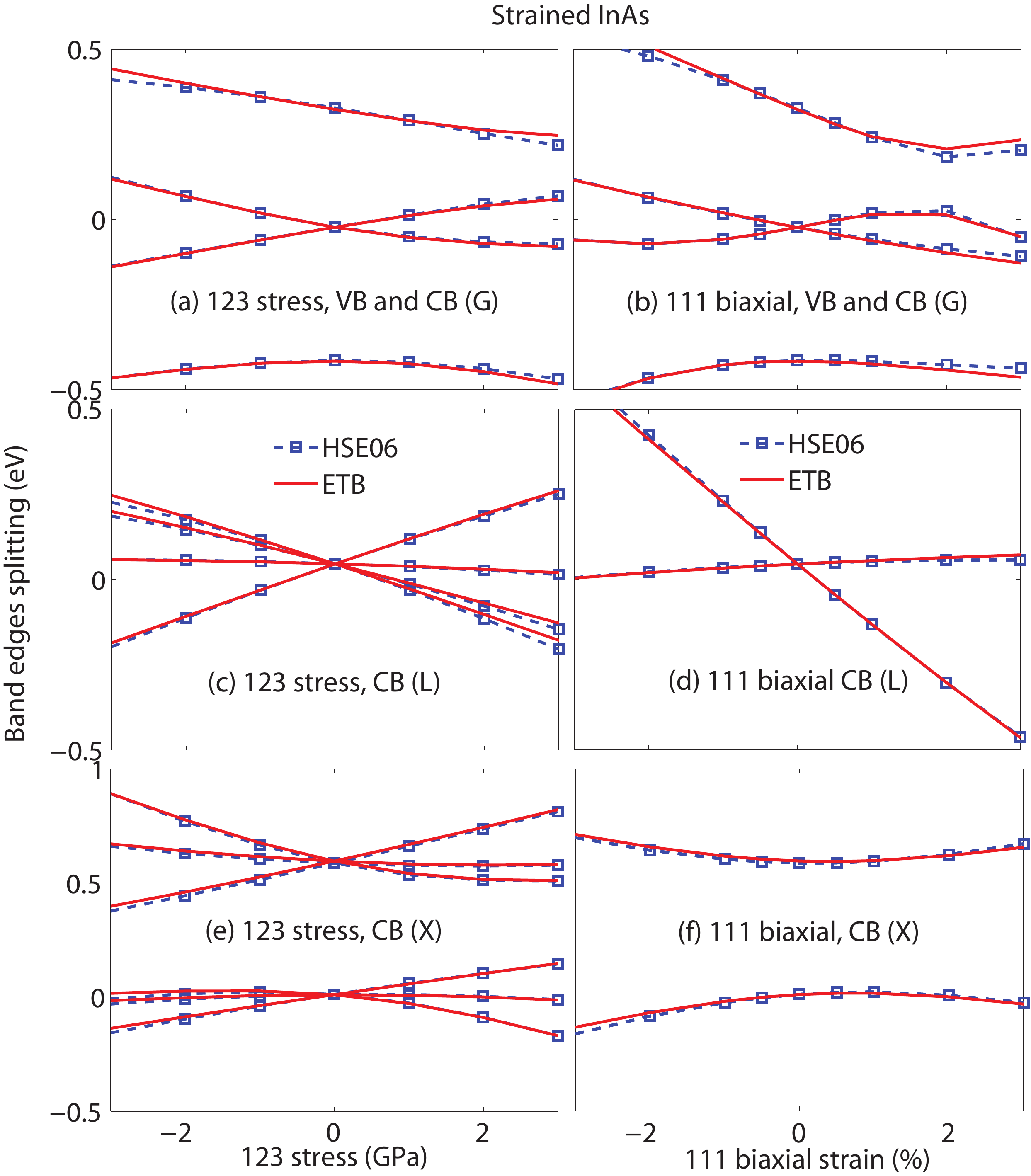}
\caption{Strain induced band edge splitting of selected conduction bands and
valence bands at $\Gamma$, $X$ and $L$ points of InAs. At $\Gamma$
point, 6 top most valence bands and 2 lowest conduction bands are
shown. 4 lowest conduction bands at $X$ points are shown. The lowest
conduction band at $L$ points are included in the figures. The
valence bands at $X$ and $L$ points are not shown as those points
are of low energy. The ETB band edge splitting are in good agreement
with the corresponding HSE06 results. }\label{fig_strain_splitting}
\end{figure}
The atom type dependent onsite parameters are listed on table
\ref{tab:onsite_soc}. Table \ref{tab: nb to onsite, GeSi/IIIVs} and
\ref{tab:interatomic_parameters_bond_length_SiGe IIIVs} summarizes
the bond length dependent onsite and interatomic coupling parameters
respectively. From table
\ref{tab:interatomic_parameters_bond_length_SiGe IIIVs}, it can be
seen that interatomic parameters for different III-V materials have
similar values. Multipole dependent onsite parameters and
interatomic coupling parameters are listed in table
\ref{tab:off_diagonal_onsite_parameters_multipole} and
\ref{tab:interatomic_coupling_parameters_multipole} respectively.
The relative band offsets are incorporated in the ETB parameters. 
The top valence bands obtained by the ETB model corresponding to the value from HSE06 calculations instead of zero. However we shifted top valence bands to zero in presented figures when showing band structures in order to improve the readability. The parameters $P$'s $Q$'s and $S$'s  in principle contain the same
number of parameters as interatomic interaction parameter $V$.
However, it turns out that it is sufficient to consider only  $s-p$,
$s-d$, $p-p$ and $d-d$ interactions for parameters $P$'s, $Q$'s and
$S$'s. Others such as $s^*-p$, $s^*-d$ and $p-d$ interactions are
constrained to zero.

\begin{table*}
    \small
\setlength{\tabcolsep}{1pt}
\begin{tabular}
[c]{c|c|c|c}\hline\hline
& AlP & GaP & InP \\\hline%
\begin{tabular}
[c]{c}%
targets \\
$b_v$\\
$d_v$\\
$\Xi_{001}$\\
$\Xi_{110}$\\
\end{tabular}
&
\begin{tabular}
[c]{cc}%
\begin{tabular}
[c]{ccc}%
HSE06 & ETB & $\;$error$\;$ \\
              $1.75$ &               $1.68$ &              $3.5\%$ \\ 
              $5.37$ &               $5.57$ &              $3.7\%$ \\ 
              $5.45$ &               $5.13$ &              $6.0\%$ \\ 
             $15.44$ &              $16.79$ &              $8.8\%$ \\
\end{tabular}
& 
\begin{tabular}
[c]{c}%
Ref \\              
              $1.5$ \\ 
              $4.6$ \\ 
              $-$ \\ 
             $-$  \\ 
\end{tabular}
\\
\end{tabular}
&
\begin{tabular}
[c]{cc}%
\begin{tabular}
[c]{ccc}
HSE06 & ETB & $\;$error$\;$\\
              $2.06$ &               $2.02$ &              $1.7\%$ \\ 
              $5.25$ &               $5.43$ &              $3.6\%$ \\ 
              $7.14$ &               $7.12$ &              $0.4\%$ \\ 
             $17.66$ &              $17.90$ &              $1.4\%$ \\
\end{tabular}
& 
\begin{tabular}
[c]{c}%
Ref \\              
              $2.0$ \\ 
              $5.0$ \\ 
              $-$ \\ 
             $-$  \\ 
\end{tabular}
\\
\end{tabular}
&
\begin{tabular}
[c]{cc}%
\begin{tabular}
[c]{ccc}
HSE06 & ETB & $\;$error$\;$ \\
              $1.72$ &               $1.63$ &              $5.1\%$ \\ 
              $4.43$ &               $4.81$ &              $8.6\%$ \\ 
              $5.64$ &               $5.55$ &              $1.5\%$ \\ 
             $17.34$ &              $18.33$ &              $5.7\%$ \\
\end{tabular}
& 
\begin{tabular}
[c]{c}%
Ref \\             
              $1.5$ \\ 
              $4.6$ \\ 
              $-$ \\ 
             $-$  \\ 
\end{tabular}
\\
\end{tabular}
\\\hline
    & AlAs & GaAs & InAs \\\hline%
\begin{tabular}
[c]{c}%
targets \\
$b_v$\\
$d_v$\\
$\Xi_{001}$\\
$\Xi_{110}$\\
\end{tabular}
&
\begin{tabular}
[c]{cc}%
\begin{tabular}
[c]{ccc}%
HSE06 & ETB & $\;$error$\;$ \\
              $1.79$ &               $1.79$ &              $0.2\%$ \\ 
              $5.47$ &               $5.81$ &              $6.3\%$ \\ 
              $5.10$ &               $4.89$ &              $4.1\%$ \\ 
             $15.57$ &              $15.21$ &              $2.3\%$ \\            
\end{tabular}
& 
\begin{tabular}
[c]{c}%
Ref \\             
              $2.3$ \\ 
              $3.4$ \\ 
              $-$ \\ 
             $-$  \\ 
\end{tabular}
\\
\end{tabular}
&
\begin{tabular}
[c]{cc}%
\begin{tabular}
[c]{ccc}
HSE06 & ETB & $\;$error$\;$ \\
              $2.11$ &               $2.00$ &              $5.5\%$ \\ 
              $5.41$ &               $5.19$ &              $4.1\%$ \\ 
              $6.55$ &               $6.62$ &              $1.1\%$ \\ 
             $17.52$ &              $17.31$ &              $1.2\%$ \\ 
\end{tabular}
& 
\begin{tabular}
[c]{c}%
Ref \\             
              $2.0$ \\ 
              $4.8$ \\ 
              $-$ \\ 
              $-$  \\ 
\end{tabular}
\\
\end{tabular}
&
\begin{tabular}
[c]{cc}%
\begin{tabular}
[c]{ccc}
HSE06 & ETB & $\;$error$\;$ \\
              $1.75$ &               $1.70$ &              $2.7\%$ \\ 
              $4.44$ &               $4.57$ &              $2.9\%$ \\ 
              $4.93$ &               $4.92$ &              $0.1\%$ \\ 
             $16.63$ &              $15.95$ &              $4.1\%$ \\ 
\end{tabular}
& 
\begin{tabular}
[c]{c}%
Ref \\             
              $1.8$ \\ 
              $3.6$ \\ 
              $-$ \\ 
              $-$  \\ 
\end{tabular}
\\
\end{tabular}
\\\hline
& AlSb & GaSb & InSb \\\hline%
\begin{tabular}
[c]{c}%
targets \\
$b_v$\\
$d_v$\\
$\Xi_{001}$\\
$\Xi_{110}$\\
\end{tabular}
&
\begin{tabular}
[c]{cc}%
\begin{tabular}
[c]{ccc}%
HSE06 & ETB & $\;$error$\;$ \\
              $1.82$ &               $1.95$ &              $6.8\%$ \\ 
              $5.44$ &               $5.20$ &              $4.3\%$ \\ 
              $5.30$ &               $5.21$ &              $1.8\%$ \\ 
             $13.96$ &              $13.42$ &              $3.8\%$ \\
\end{tabular}
& 
\begin{tabular}
[c]{c}%
Ref \\             
              $1.35$ \\ 
              $4.3$ \\ 
              $-$ \\ 
              $-$  \\ 
\end{tabular}
\\
\end{tabular}
&
\begin{tabular}
[c]{cc}%
\begin{tabular}
[c]{ccc}
HSE06 & ETB & $\;$error$\;$ \\
              $2.14$ &               $2.27$ &              $6.0\%$ \\ 
              $5.32$ &               $5.38$ &              $1.1\%$ \\ 
              $8.14$ &               $7.85$ &              $3.6\%$ \\ 
             $15.32$ &              $14.29$ &              $6.7\%$ \\
\end{tabular}
& 
\begin{tabular}
[c]{c}%
Ref \\             
              $2.0$ \\ 
              $4.7$ \\ 
              $-$ \\ 
              $-$  \\ 
\end{tabular}
\\
\end{tabular}
&
\begin{tabular}
[c]{cc}%
\begin{tabular}
[c]{ccc}
HSE06 & ETB & $\;$error$\;$ \\
              $1.80$ &               $1.89$ &              $5.2\%$ \\ 
              $4.60$ &               $4.67$ &              $1.5\%$ \\ 
              $7.60$ &               $7.48$ &              $1.5\%$ \\ 
             $14.57$ &              $14.11$ &              $3.2\%$ \\             
\end{tabular}
& 
\begin{tabular}
[c]{c}%
Ref \\             
              $2.0$ \\ 
              $4.7$ \\ 
              $-$ \\ 
              $-$  \\ 
\end{tabular}
\\
\end{tabular}
\\\hline
\end{tabular}
\centering \caption{Targets comparison of deformation potentials of
III-V materials.  The Reference bandedge and effective masses are from Ref \onlinecite{IIIV_para_Vurgulfman}.
}%
\label{tab:strained_targets_comparison_IIIV}
\end{table*}

\begin{figure}\centering
\includegraphics[width=0.4\textwidth]{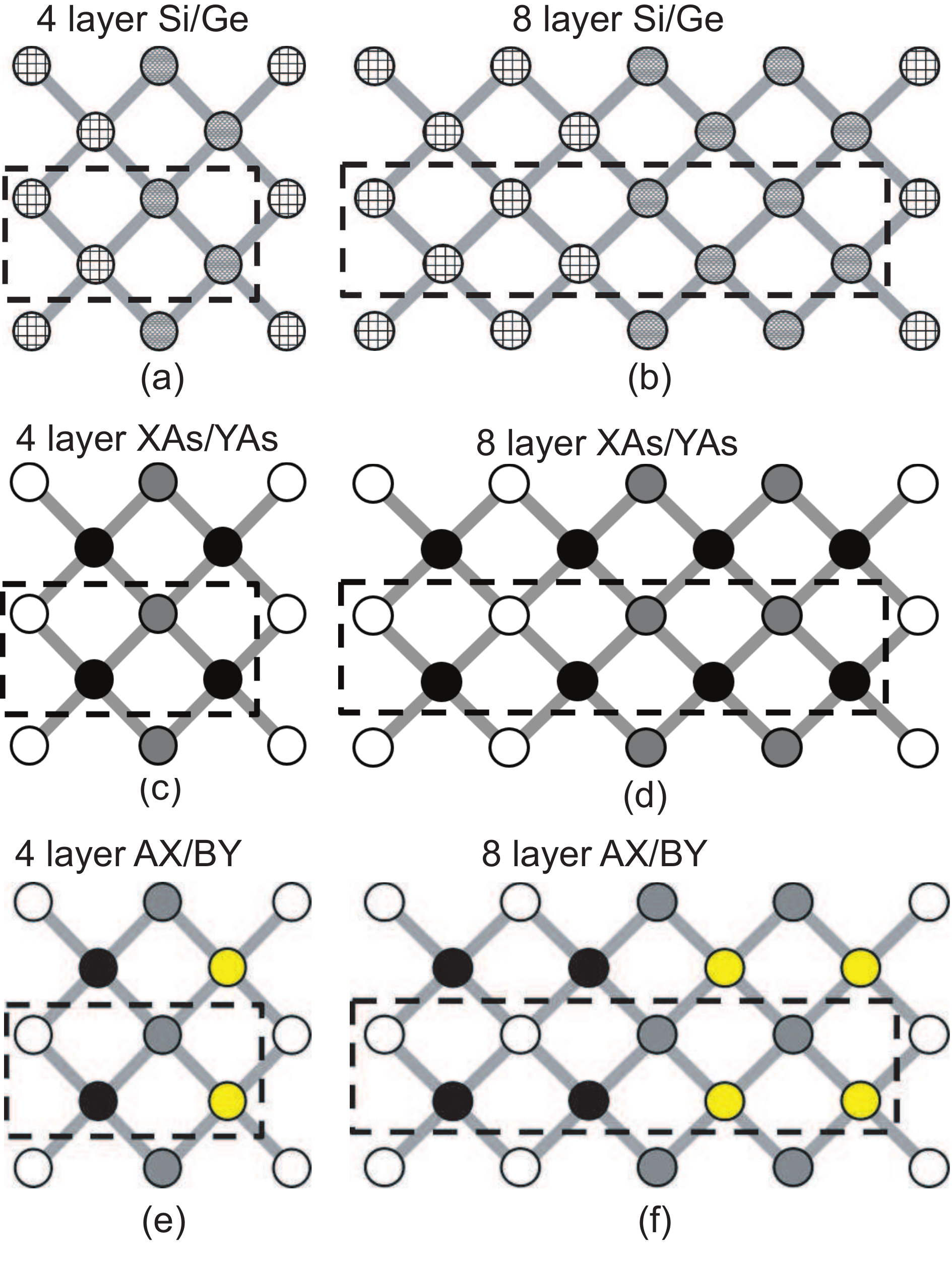}
\caption{Atom structure of Si/Ge and XAs/YAs type superlattices. (a)
Si/Ge superlattice with 4 layers in the unit cell; (b)  Si/Ge
superlattice with 8 layers in the unit cell. (c)  XAs/YAs
superlattice with 4 layers in the unit cell; (d)  XAs/YAs
superlattice with 8 layers in the unit cell. (e)  AX/BY superlattice
with 4 atoms in the unit cell; (f)  AX/BY superlattice with 8 layers
in the unit cell. The primitive unit cells are marked by dashed
lines. } \label{fig_superlattice_structures}
\end{figure}

\subsection{Unstrained and strained materials}\label{sec:strained_materials_cmp}
Fig. \ref{fig_IVs} and \ref{fig_IIIVs} show band structures of
unstrained bulk band structure for group IV and III-V materials. The
presented materials include Si, Ga, XP, XAs and XSb with X =
Al,Ga,In. It can be seen that the ETB results of unstrained bulk
group IV and III-V  materials match corresponding HSE06 results
well. Tables
\ref{tab:targets_comparison_SiGe},\ref{tab:targets_comparison_XP},\ref{tab:targets_comparison_XAs}
and \ref{tab:targets_comparison_XSb} compare the effective masses
and critical band edges between ETB and HSE06 calculations.
Most of the effective masses of important valence and conduction valleys are within 10\% error. Effective masses of higher conduction valleys like $m_l$ or L valleys tend to have larger error.
Discrepancies of critical band edges at high
symmetric points between ETB and HSE06 are within 10meV.

Fig.\ref{fig_Si_hydro_static} shows Si band structures under
hydrostatic strain. The hydrostatic strain does not change crystal
symmetry, thus the degeneracy at high symmetry points conserve under
hydro static strain. However, it can be observed by comparing
Fig.\ref{fig_Si_hydro_static} (a) and (b) that the hydrostatic
strains change the band edges significantly. With a lattice constant
of $5.4\AA$, the lowest conduction bands of Si are $X$ valleys, the
$L$ and s-type $\Gamma$ valley (Ecs(G)) are of more than 1eV above
the $X$ valleys. However, with a larger lattice constant of
$5.8\AA$, the $L$ and $\Gamma$ gap descend dramatically , while the
$X$ gap even increase slightly. The change of band gaps are shown
clearly by Fig.\ref{fig_Si_hydro_static} (c), it can be seen that at
around $5.8\AA$, the $L$ and s-type $\Gamma$ valley become lower
than the $X$ valleys. As the lattice constant increase more, Si
becomes a direct gap material (lowest conduction band is $\Gamma$
valley). In fact, if the lattice constant is sufficiently large, Si
becomes a metal as the s-type $\Gamma$ valley conduction band become
even lower than the valence bands. The trend shown by
Fig.\ref{fig_Si_hydro_static} is valid for other group IV and III-V
materials which have diamond or zincblende structures.

Fig. \ref{fig_strain_splitting} shows the band edge splitting at
$\Gamma$, $X$ and $L$ points of InAs under different strains (strain produced by uniaxial stress along [123] direction and biaxial strain along [111]). The strain presented were not considered in the fitting process and produces complicated bandedge splitting especially for X and L valleys. It can be seen that the ETB band edge splittings are in good agreement with
the corresponding HSE06 results. To quantitatively estimate the
discrepancies between ETB and HSE06 calculations for strained
materials, the deformation potentials are extracted from both ETB
and HSE06 results. The deformation potentials of group IV and III-V
materials are compared in tables
\ref{tab:strained_targets_comparison_Ge_Si} and
\ref{tab:strained_targets_comparison_IIIV}. It can be seen that the
important deformation potentials by ETB agree well with the HSE06
results. The discrepancies are within 2\%. The deformation
potentials $b_v$ and $d_v$ describe the band edge splitting of
valence bands under diagonal and off-diagonal strain components
respectively. $\Xi_{001}$ and $\Xi_{110}$ describe the conduction
band edge splitting at X points due to diagonal and off-diagonal
strain components respectively. The definition of those deformation
potentials are specified in Appendix \ref{app:definition of
deformation potentials}.

\begin{figure*}
\includegraphics[width=0.8\textwidth]{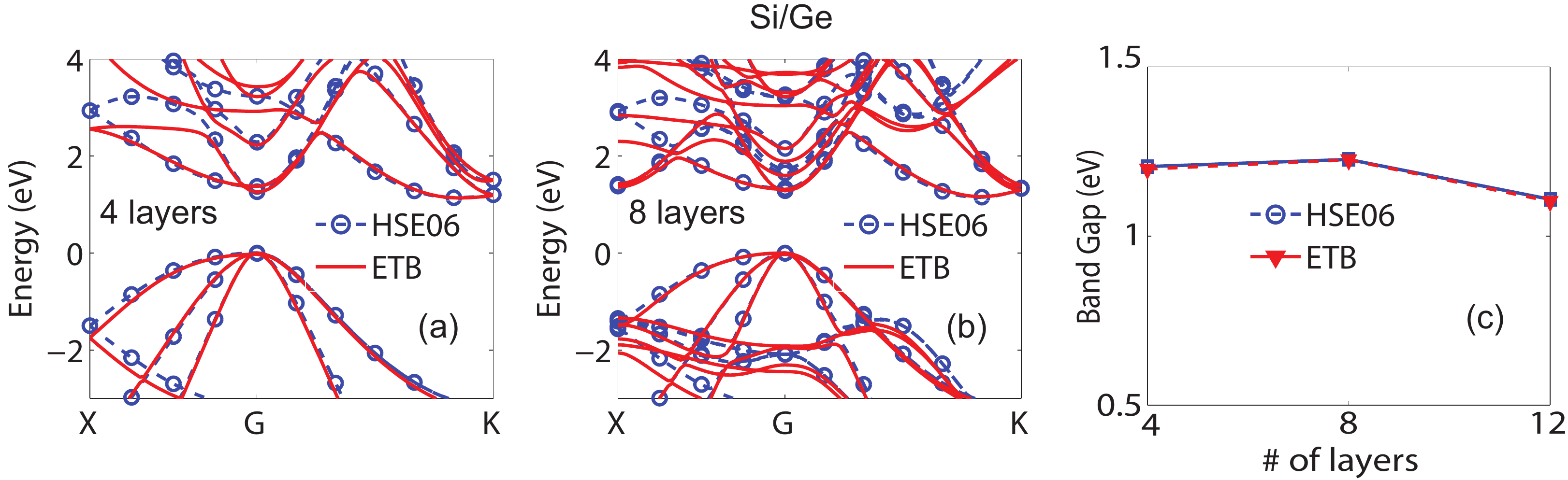}
\caption{Band structures of Si/Ge superlattices by ETB and HSE06
calculations. Figures correspond to supercells which contain 4 atoms
(a) and 8 atoms (b) and band gaps of Si/Ge superlattices verse
number of atoms in the supercell (c). }
\label{fig_GeSi_superlattices}
\end{figure*}
\begin{figure*}
\includegraphics[width=0.8\textwidth]{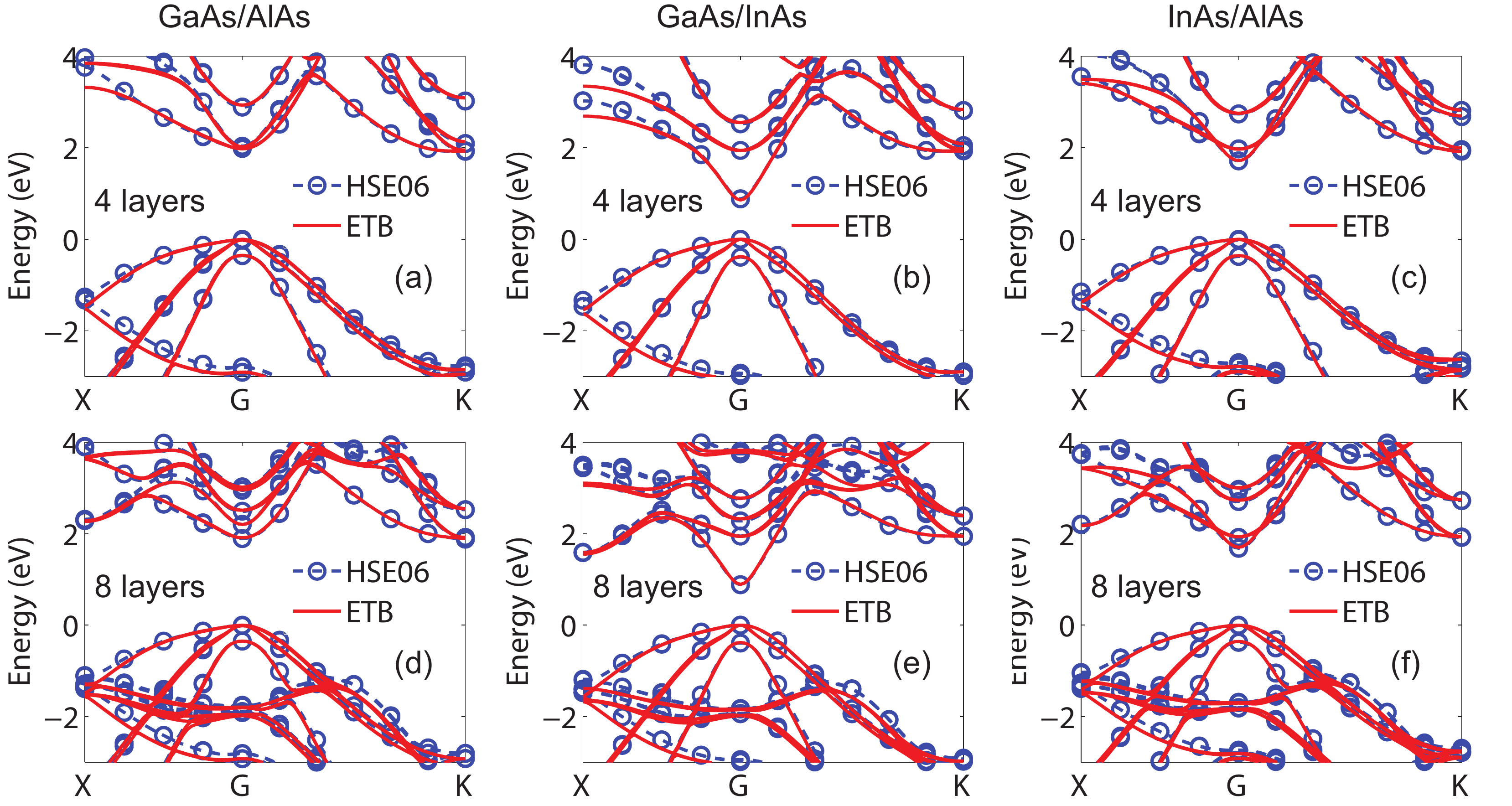}
\caption{Band structures of Arsenides superlattices by ETB and HSE06
calculations. Presented band structures include band structures of
superlattices of 001 AlAs/GaAs((a),(d)), InAs/GaAs((b),(e)) and
InAs/AlAs((c),(f)). Upper figures correspond to supercells which
contain 4 atoms(Fig \ref{fig_superlattice_structures} (a)), while
lower figures corresponds to supercells with 8 atoms (Fig
\ref{fig_superlattice_structures} (b)). }
\label{fig_XAs_superlattices}
\end{figure*}
\subsection{Tight binding analysis of superlattices }\label{sec:superlattices_cmp}
To investigate the transferability of our ETB parameters,band structures of  group IV and group III-V superlattices are calculated by both ETB and HSE06
models. The atom structures of the superlattices considered in this work
are shown in Fig.\ref{fig_superlattice_structures}. The
superlattices considered in this work grow along 001 direction.
Those superlattices contain only a few layers of atoms (with thickness
from about 0.5 nm to 1.5 nm). To model those superlattices by ETB
method, in principle, self-consistent ETB calculations with Possion
equation should be applied if there is charge redistribution in the
hetero-structures. However the presented superlattices turn out to
be either type I or type II heterojunctions as the
\textit{ab-initio} band structures shows band gap of at least 0.5eV
for all the presented superlattices. The charge redistribution in
type I or II heterostructures under zero temperature is negligible
because the valence bands of both materials are perfectly occupied.
The negligible build-in field can also be realized by looking at the
envelope of \textit{ab-initio} local
potentials\cite{VandeWalle_modelsolid_IIIV,VandeWalle_modelsolid_SiGe}.
Thus, the presented ETB calculations for superlattices all assumes
zero build-in potentials. The parameter $\delta d_{ij}$ are all set
to zero in order to compare with \textit{ab-initio} results.


\begin{figure*}\centering
\includegraphics[width=0.8\textwidth]{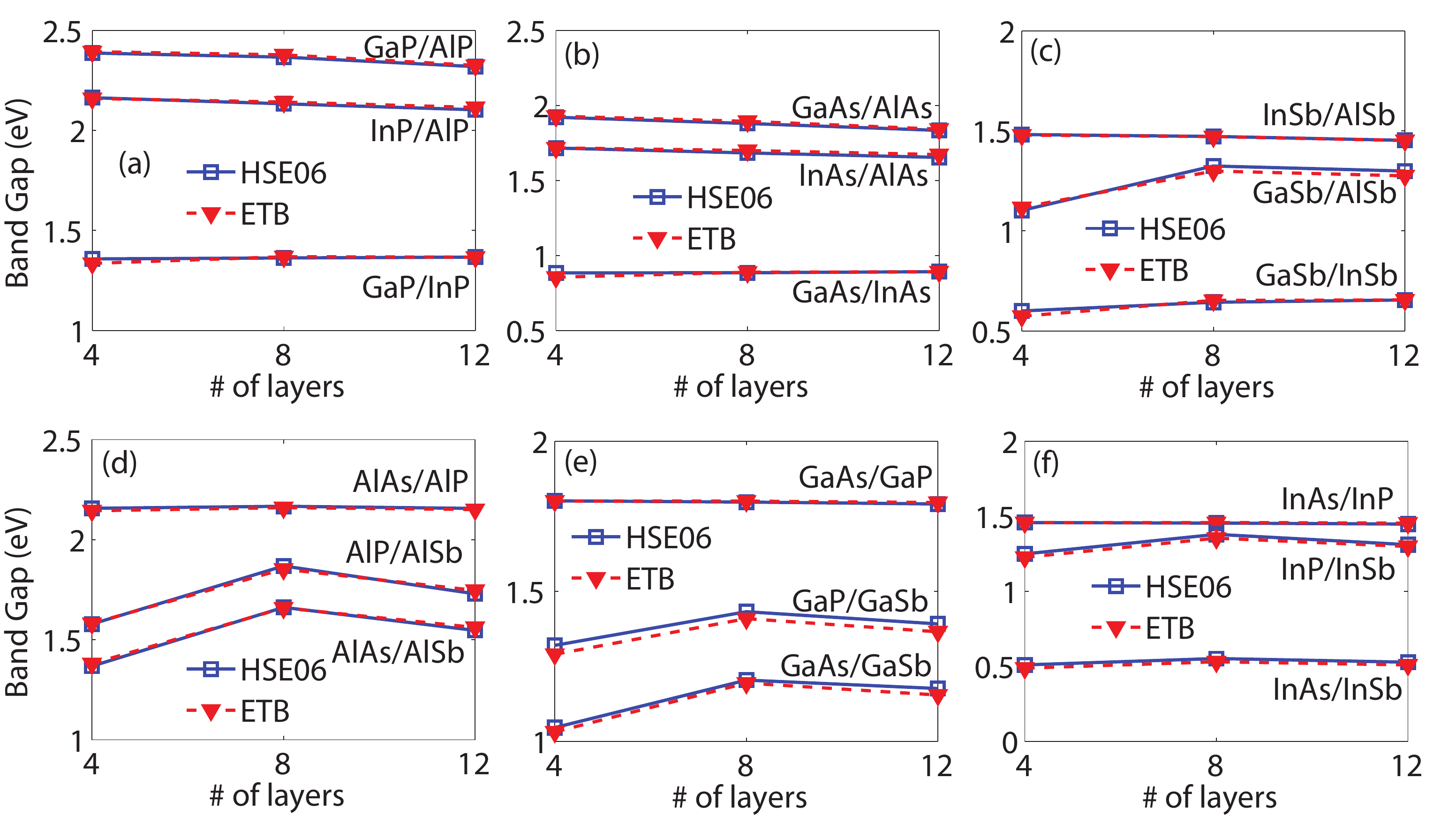}
\caption{Band gaps of III-V superlattices by ETB and HSE06
calculations. The presented band gaps include superlattices of (a)
XP/YP , (b) XAs/YAs and (c) XSb/YSb with ( X and Y stand for
different cations, X,Y = Al, Ga or In) and (e) AlX/AlY , (f) GaX/GaY
and (g) InX/InY with ( X and Y stand for different anions, X,Y = P,
As or Sb). The ETB band gaps of different superlattices show good
agreement with HSE06 results, demonstrating the ETB parameters have
good transferability.} \label{fig_superlattices_band_gap}
\end{figure*}
\begin{figure*}
\includegraphics[width=0.8\textwidth]{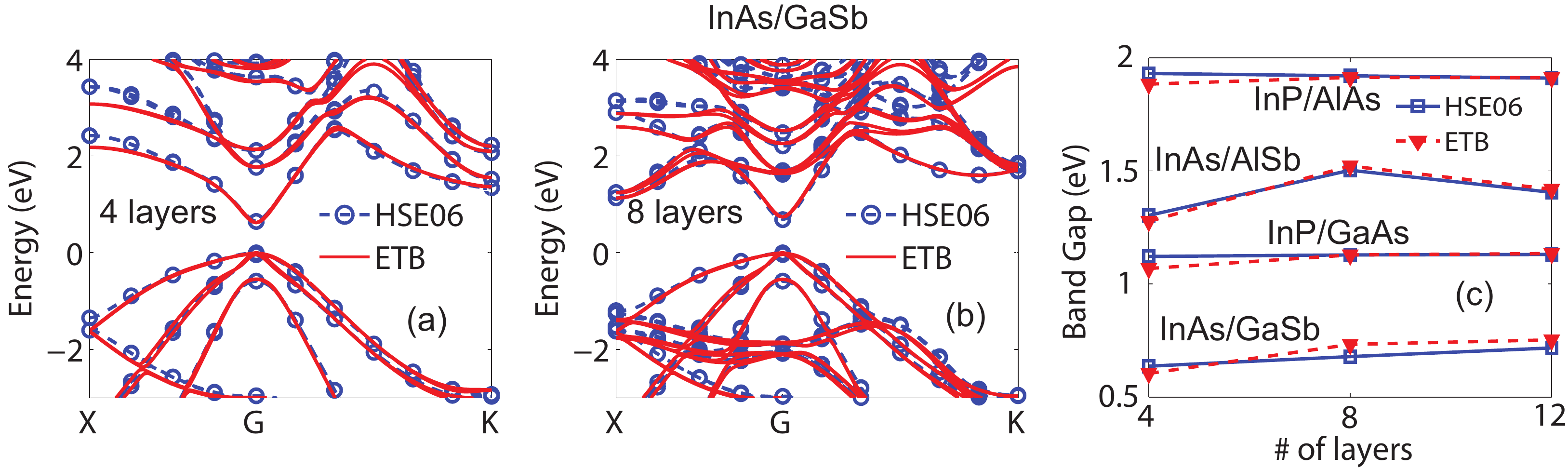}
\caption{Band structures InAs/GaSb superlattices by ETB and HSE06
calculations. Present figures include band structures of 4 layer (a)
and 8 layer (b) InAs/GaSb superlattices. Band gaps of AX/BY type
superlattices are shown in (c); InAs/GaSb, InAs/AlSb, InP/GaAs and
InP/AlAs superlattices are considered. }
\label{fig_InAsGaSb_superlattices}
\end{figure*}

Fig. \ref{fig_GeSi_superlattices} and Fig.
\ref{fig_XAs_superlattices} show the comparison of band structures
of Si/Ge and Arsenides superlattices by ETB and Hybrid functional
calculations respectively. In these figures, band structures of
Si/Ge, GaAs/AlAs,
GaAs/InAs and InAs/AlAs superlattices are presented. 
In both ETB and hybrid functional calculations, zero temperature is
assumed. For each type of superlattices, band structure of two
different unit cells are shown. It can be seen that the ETB band
structures are in good agreement for energy from -2eV to 1eV above
lowest conduction bands. ETB band structures are obtained with the
parameters given by previous sections without introducing extra
fitting parameters. From Fig. \ref{fig_GeSi_superlattices} and Fig.
\ref{fig_XAs_superlattices}, it can be seen that ETB calculations
without solving Poisson equation (zero build-in potential is added )
match the HSE06 results well. More complicated cases include
InAs/GaSb superlattices which contain no common cations or anions at
material interface. The InAs/GaSb superlattices with 4 atomic layers
can also be interpreted as InSb/GaAs superlattice. From Fig.
\ref{fig_InAsGaSb_superlattices} (a) and (b), it can be seen that
ETB calculations match the HSE06 results well even for interfaces
with no common cations or anions.

In 001 superlattices, the primitive unit cells are defined by
vectors $\mathbf{u}_1 = [0.5,0.5,0]a, \mathbf{u}_2 = [-0.5,0.5,0]a $
and $\mathbf{u}_3 = [0,0,N]a$, where $N$ can be any integer number.
According to the theory of Brillouin zone
folding\cite{Boykin_BZ_unfolding,Boykin_BZ_unfolding_alloy,Boykin_BZ_unfolding_SiGe,Unfolding_Zunger},
the $X$ points along [001] direction in a fcc Brillouin zone is
folded to the $k = [0,0,0]$ point in the Brillouin zone of
superlattices. As a result, the lowest few conduction states at $k =
[0,0,0]$ of 001 superlattices can have the feature of $\Gamma$ and
$X$ conduction valleys in pure materials. The $\Gamma$ and $X$
conduction valleys can be easily distinguished by the corresponding
ETB wave functions. Considering the valleys in a fcc Brillouin zone,
the lowest conduction states at $\Gamma$ point are dominated by s
and s* orbitals; while the conduction states at $X$ points have
significant contribution from both s and p orbitals. This can also
be realized by the effective masses of the valleys. The folded X
conduction valleys have anisotropic effective masses as it is shown
in Fig.\ref{fig_XAs_superlattices} (a) and (d); while the $\Gamma$
valley have isotropic effective masses as in
Fig.\ref{fig_XAs_superlattices} (b) and (e).  It can be seen from
Fig.\ref{fig_XAs_superlattices} that the lowest conduction state in
AlAs/GaAs superlattices have the feature of X conduction valley;
while in InAs/GaAs and InAs/AlAs superlattices, the lowest
conduction state has the feature of $\Gamma$ valley.

Fig. \ref{fig_GeSi_superlattices} (c),
Fig.\ref{fig_superlattices_band_gap} and
Fig.\ref{fig_InAsGaSb_superlattices} (c) compare the ETB band gap of
for different superlattices with corresponding HSE06 results. Fig.
\ref{fig_GeSi_superlattices} (c) shows the band gaps in Si/Ge
superlattices. The compared superlattices in
Fig.\ref{fig_superlattices_band_gap} include superlattices with
common anions (XP/YP, XAs/YAs and XSb/YSb) and superlattices with
common cations (AlX/AlY, GaX/GaY and InX/InY).
Fig.\ref{fig_InAsGaSb_superlattices} (c) shows the band gaps of
selected AX/BY type superlattices, including InAs/GaSb, InAs/AlSb,
InP/GaAs and InP/AlAs. For the superlattices shown in the figure,
averaged lattice constant is used to create the unit cell of the
superlattices since lattice mismatch always exists in superlattices.
It can be seen that ETB methods in this work delivered accurate band
gaps for ultra small superlattices.  For ultra small superlattices,
the band gaps are not always monotonic functions of thickness. This
non-monotonic dependency of band gaps can be seen in many of the
presented superlattices which have common cations
(Fig.\ref{fig_superlattices_band_gap} (d), (e) and (f)). The ETB
band gap of superlattices agree well with corresponding HSE06. For superlattices which contain common cations or anions (shown in Fig.\ref{fig_superlattices_band_gap}), the
largest discrepancy of about 0.03eV appears in GaP/GaSb superlattices.
While the discrepancy of superlattices which contain no common cation or anions, the largest discrepancy reaches a slightly higher of about 0.05eV.
These comparisons suggest that the ETB model and parameters by this
work has good transferability.

\section{Conclusion}\label{sec:Conclusion}
Environment dependent ETB model with nearest neighbor interactions is developed. ETB parameters for group IV and III-V semiconductors are parameterized with respect to
HSE06 calculations. Good agreements are achieve for unstrained and
arbitrarily strained materials. The ETB parameters show good transferability
when applied to ultra-small superlattices. The ETB band structures of
superlattices match the corresponding HSE06 result well. Tight
binding band gaps of varieties of superlattices show less than 0.1
eV discrepancies compared with HSE06 calculations.
This work demonstrated that an ETB model with good transferability 
can be achieved with nearest neighbor interactions for group IV and III-V materials.

\begin{acknowledgments}
The use of nanoHUB.org computational resources operated by the
Network for Computational Nanotechnology funded by the US National
Science Foundation under Grant Nos. EEC-0228390, EEC-1227110,
 EEC-0634750, OCI-0438246, OCI-0832623 and OCI-0721680
is gratefully acknowledged. Samik Mukherjee and Evan Wilson from Network
for Computational Nanotechnology, Purdue University are
acknowledged for helpful discussion and suggestions.
\end{acknowledgments}

\appendix

\section{Expression of $\mathcal{M}^{(l)}_{\alpha,\gamma}\left(\mathbf{\hat{d}}\right)$}\label{app:D_matrix}
For a unit vector $\mathbf{\hat{d}} = [x,y,z]$,
the explicit form of $\mathcal
{M}^{(l)}_{\alpha,\gamma}\left(\mathbf{\hat{d}}\right)$ are given as
follows. For p and d orbitals, the order of orbitals are arranged
according to quantum number $m$, with $\{p_y,p_z,p_x\}$ and
$\{d_{xy},d_{yz},d_{2z^2-x^2-y^2},d_{xz},d_{x^2-y^2}\}$.
Here the $\mathcal
{M}^{(l)}_{\alpha,\gamma}\left(\mathbf{\hat{d}}\right)$ are written as matrices with $\alpha$ and $\gamma$ as row and column indices respectively. 
 
The matrix  $[\mathcal
{M}^{(1)}_{00,1m'}\left(\mathbf{\hat{d}}\right)]$ is given by
\begin{equation}\label{eq:D_011}
    \frac{\sqrt{3}}{4\pi}\left[
      \begin{array}{ccc}
        y & z & x \\
      \end{array}
    \right].
\end{equation}

 The matrix  $[\mathcal
{M}^{(1)}_{1m,2m'}\left(\mathbf{\hat{d}}\right)]$ is given by
\begin{equation}\label{eq:D_112}
    \frac{\sqrt{3}}{4\sqrt{5}\pi}\left[
      \begin{array}{ccccc}
      \sqrt{3}x &  \sqrt{3}z & - y &  0 & -\sqrt{3}y \\
      0 & \sqrt{3}y &  2z & \sqrt{3}x & 0 \\
      \sqrt{3}y & 0 & -x & \sqrt{3}z & \sqrt{3}x \\
      \end{array}
    \right].
\end{equation}

 The matrix $[\mathcal
{M}^{(2)}_{1m,1m'}\left(\mathbf{\hat{d}}\right)]$ is given by
\begin{equation}\label{eq:D_121}
    \frac{3}{4\pi}\left[
      \begin{array}{ccc}
        \frac{2y^2-x^2-z^2}{3} & yz & yx \\
        yz & \frac{2z^2-x^2-y^2}{3} & xz \\
        yx & xz & \frac{2x^2-y^2-z^2}{3} \\
      \end{array}
    \right].
\end{equation}

The matrix $[\mathcal
{M}^{(2)}_{2m,2m'}\left(\mathbf{\hat{d}}\right)]$ is given by
\begin{widetext}
\begin{equation}\label{eq:D_222}
    \mathcal {M}^{(2)}_{m,m'}\left(\mathbf{\hat{d}}\right)=
    \frac{15}{28\pi}
    \left[
    \begin{array}{ccccc}
     -\frac{2z^2-x^2-y^2}{3}   & xz                    & -\frac{2}{\sqrt{3}} xy    &  yz                   &  0 \\
        \hat{x}z                     &-\frac{2x^2-y^2-z^2}{3}& \frac{1}{\sqrt{3}}yz     &  xy                   & -yz \\
        -\frac{2}{\sqrt{3}} xy & \frac{1}{\sqrt{3}}yz  & \frac{2z^2-x^2-y^2}{3}    & \frac{1}{\sqrt{3}}yz  & -\frac{ x^2-y^2 }{\sqrt{3}}\\
        yz                     & xy                    &  \frac{1}{\sqrt{3}}xz     &-\frac{2y^2-x^2-z^2}{3}& xz \\
        0                      & -yz                   &-\frac{ x^2-y^2 }{\sqrt{3}}& xz                    &  -\frac{2z^2-x^2-y^2}{3} \\
    \end{array}
    \right].
\end{equation}
\end{widetext}

\section{Dipole potentials}\label{app:prove_approximation_dipole}
The interatomic coupling due to multipole was given by equation
(\ref{eq:interatomic_coupling_multipole}). For dipole moment, the
term $\mathcal{M}^{(1)}_{\alpha,\gamma}(\mathbf{\hat{d}})$ are given
by equations (\ref{eq:D_011}) and (\ref{eq:D_112}). The explicit
form of $V^{(1)}_{\alpha,\beta}$ are given in this appendix. For
example,
the ${p_x}-p_{y}$ couplings $V^{(1)}_{x,y}$ is given by
\begin{eqnarray}\label{eq:interatomic_coupling_multipole_xy}
  V^{(1)}_{x,y} = &  \sum_{k} \mathcal{M}^{(1)}_{x,s}(\mathbf{\hat{d}}_{ik}) Q^{(1)}_{s,y}(d_{ik}) \\
     & + \sum_{k'}Q^{(1)}_{x,s}(d_{jk'})
     \mathcal{M}^{(1)}_{s,y}(\mathbf{\hat{d}}_{jk'})\nonumber\\
     & +
     \sum_{d,k'}\mathcal{M}^{(1)}_{x,d}(\mathbf{\hat{d}}_{ik})Q^{(1)}_{d,y}(d_{ik})\nonumber\\
     & + \sum_{d,k} Q^{(1)}_{x,d}(d_{jk'})
     \mathcal{M}^{(1)}_{d,y}(\mathbf{\hat{d}}_{jk'})
 \nonumber,
\end{eqnarray}
The $Q^{(1)}_{\alpha,\beta}$'s are two center integrals given by
equation (\ref{eq:interatomic_coupling_Q}). Using the explicit
expression of $\mathcal{M}$ and Slater Koster formula of $Q^{(1)}$,
the terms in equation (\ref{eq:interatomic_coupling_multipole_xy})
are written as
\begin{eqnarray}\label{eq:interatomic_coupling_multipole_term1}
\sum_{k} \mathcal{M}^{(1)}_{x,s}(\mathbf{\hat{d}}_{ik}) Q^{(1)}_{s,y}(d_{ik}) & = \sum_{k}x_{ij}y_{ik}  Q^{(1)}_{sp\sigma}(d_{ik})\\
\sum_{k'}Q^{(1)}_{x,s}(d_{jk'})\mathcal{M}^{(1)}_{s,y}(\mathbf{\hat{d}}_{jk'})
& = \sum_{k'}x_{ij}y_{jk'}  Q^{(1)}_{sp\sigma}(d_{jk'})
\nonumber,
\end{eqnarray}
\begin{widetext}
\begin{eqnarray}\label{eq:interatomic_coupling_multipole_term2}
\sum_{d,k'}\mathcal{M}^{(1)}_{x,d}(\mathbf{\hat{d}}_{ik})Q^{(1)}_{d,y}(d_{ik})
 & =  \frac{1}{\sqrt{15}} x_{ij}y_{ij}p_{ij,k} \left( 3Q^{(1)}_{pd\sigma}(d_{ik}) - 2\sqrt{3}Q^{(1)}_{pd\pi}(d_{ik})\right)+x_{ik}y_{ij} \left(-Q^{(1)}_{pd\sigma}(d_{ik}) + 3\sqrt{3}Q^{(1)}_{pd\pi}(d_{ik})  \right) \\
\sum_{d,k} Q^{(1)}_{x,d}(d_{jk'})
\mathcal{M}^{(1)}_{d,y}(\mathbf{\mathbf{\hat{d}}}_{jk'}) & =
\frac{1}{\sqrt{15}} x_{ij}y_{ij}p_{ij,k'} \left(
3Q^{(1)}_{pd\sigma}(d_{jk'}) -
2\sqrt{3}Q^{(1)}_{pd\pi}(d_{jk'})\right) - x_{jk'}y_{ij}
\left(-Q^{(1)}_{pd\sigma}(d_{jk'}) +
3\sqrt{3}Q^{(1)}_{pd\pi}(d_{jk'})  \right)
 \nonumber,
\end{eqnarray}
\end{widetext} The $p_{ij,k} = \sum_{m} Y_{1,m}(\Omega_{\hat{d}_{ij}})Y_{1,m}(\Omega_{\hat{d}_{ik}})$ and
$p_{ji,k'} = \sum_{m}
Y_{1,m}(\Omega_{\hat{d}_{ji}})Y_{1,m}(\Omega_{\hat{d}_{jk'}})$,
satisfying $\sum_{k} p_{ij,k} = p_{ij}$ and $\sum_{k} p_{ji,k'} =
p_{ji}$ with $p_{ij}$ and $p_{ji}$ given by equations
(\ref{eq:p_ij}) and (\ref{eq:p_ji}). It can be seen that the terms
with $p_{ij}$ or $p_{ji}$ has resemblance with Slater Koster formula
of $V_{xy} = xy\left( V_{pp\sigma} -V_{pp\pi}\right)$.
To make the expression simpler, in this work, only the terms with
$p_{ij,k}$ and $p_{ji,k'}$ are preserved. Let
\begin{eqnarray}
\frac{3Q^{(1)}_{pd\sigma}}{\sqrt{15}}(d_{ik})&=& \frac{4\pi}{3} \left(P_{pp\sigma} +
\frac{\delta \mathbf{d}_{ik}}{d_0} S_{pp\sigma}\right)  \\
\frac{2\sqrt{3}Q^{(1)}_{pd\pi}}{\sqrt{15}}(d_{ik})&=& \frac{4\pi}{3} \left( P_{pp\pi} +
\frac{\delta \mathbf{d}_{ik}}{d_0} S_{pp\pi}\right)
\end{eqnarray}
The $V^{(1)}_{x,y}$ can be approximated by
\begin{equation}\label{eq:V1_xy}
    V^{(1)}_{x,y} =
    x_{ij}y_{ij}( \delta V^{(1)}_{pp\sigma} - \delta V^{(1)}_{pp\pi} ),
\end{equation}
here the $\delta V^{(1)}_{pp\sigma}$ and $\delta V^{(1)}_{pp\pi}$ are defined by
\begin{eqnarray}\label{eq:V1_l}
    \delta V^{(1)}_{pp\sigma} & =\frac{4\pi}{3}(p_{ij}+p_{ji})P_{pp\sigma} + \frac{4\pi}{3}(q_{ij}+q_{ji})S_{pp\sigma} \\
    \delta V^{(1)}_{pp\pi} & =\frac{4\pi}{3}(p_{ij}+p_{ji})P_{pp\pi} + \frac{4\pi}{3}(q_{ij}+q_{ji})S_{pp\pi}
\end{eqnarray}
The $p_{ij}$, $p_{ji}$, $q_{ij}$ and $q_{ji}$ are given by
equations (\ref{eq:p_ij}) and (\ref{eq:p_ji}). Similar process can be applied to other $V^{1}_{\alpha,\beta}$'s. The generalized approximation was summarized by equation (\ref{eq:coupling_dipole}).

\section{deformation potential}\label{app:definition of deformation potentials}
  \begin{itemize}
    \item deformation potentials of top valence bands is defined by a 4 band Luttinger k.p
    Hamiltonian at $\Gamma$ point.
    \begin{equation}\label{eq:deformation_potential_vb}
    H_{\varepsilon} =- \left[
                        \begin{array}{cccc}
                          P_{\varepsilon}+Q_{\varepsilon} & -S_{\varepsilon} & R_{\varepsilon} & 0 \\
                          -S_\varepsilon^{\dag} & P_{\varepsilon}-Q_{\varepsilon} & 0 & R_{\varepsilon} \\
                          R^\dag & 0 & P_{\varepsilon}-Q_{\varepsilon} & S_{\varepsilon} \\
                          0 & R_{\varepsilon}^\dag & S_{\varepsilon}^{\dag} & P_{\varepsilon}+Q_{\varepsilon} \\
                        \end{array}
                      \right]
    \end{equation} with
    \begin{eqnarray}
     & P_\varepsilon &= -a_v\left( \varepsilon_{xx} + \varepsilon_{yy} + \varepsilon_{zz} \right)  \\
     & Q_\varepsilon &= -\frac{b_v}{2} \left( \varepsilon_{xx} + \varepsilon_{yy} - 2 \varepsilon_{zz} \right)  \\
     & R_\varepsilon &= \frac{\sqrt{3}b_v}{2} \left( \varepsilon_{xx} - \varepsilon_{yy} \right) -id_v
     \varepsilon_{xy}\\
     & S_\varepsilon &= -d_v(\varepsilon_{xz} - \varepsilon_{yz})
    \end{eqnarray}
    This 4 band Hamiltonian describe the strain behavior top valence bands of zincblende and diamond structures.
    $b_v$ describe the the Hole splitting under 001 strains( $\varepsilon_{xx} = \varepsilon_{yy} = -0.5\varepsilon_{zz}, \textrm{or} $).
    $d_v$ describes the Hole splitting under shear components ($\varepsilon_{xy}$, $\varepsilon_{yz}$, $\varepsilon_{xz}$).

    \item the deformation potential of CB(X valleys)\cite{PeterYu},
    \begin{equation}\label{eq:deformation_potential_CB}
        E_c = \Xi_{001} (\hat{k} \cdot \varepsilon \cdot \hat{k} )
    \end{equation}
    where $\varepsilon$ is the strain tensor, $\hat{k}$ is a unit
    vector along the direction of one of the conduction band minima.

    and the deformation potential of conduction X valleys due to $\varepsilon_{xy}$ is
    described by 2 band Hamiltonian
    \begin{equation}\label{eq:deformation_potential_CB}
        \left[
          \begin{array}{cc}
             E_{u} & \Xi_{110} \varepsilon_{xy} \\
             \Xi_{110} \varepsilon_{xy}  & E_{l} \\
          \end{array}
        \right].
    \end{equation}
  \end{itemize}
  This Hamiltonian describes the upper and lower conduction bands at $X$ point of zincblende and diamond structures.
  The energy difference $\Delta E$ between the upper and lower conduction bands has the relation
  $\Delta E = \sqrt{ \left(E_u-E_l\right)^2 + 4\Xi_{110}^2\varepsilon_{xy}^2 } $



\end{document}